\documentclass[12pt]{article}

\usepackage[french,english]{babel}
\usepackage{epsfig,latexsym,amsfonts,amsmath,amsthm,amssymb,
amsbsy,multirow,slashed,wasysym,textcomp,comment,mathtools,cancel,
cite,datetime,subcaption,mathtools,tensor,amscd,tikz,slashed,hyperref}

\usetikzlibrary{calc}   
\usetikzlibrary{topaths}
\usetikzlibrary{decorations}
\usetikzlibrary{decorations.pathmorphing}

 \newcommand{\badat}{\begin{alignedat}}
 \newcommand{\eadat}{\end{alignedat}}

\long\def\new#1\endnew{{\bf #1}}		
\long\def\del#1\enddel{}

\def\del{\partial}

\usepackage{color}
\usepackage{bm}

\newcommand{\pink}[1]{\textcolor{\pink}{#1}}

\definecolor{dblue}{rgb}{0.2,0.50,0.80} 

 \newcommand{\virg}{\hspace{1 mm}, \hspace{8 mm}}
\newcommand{\p}{\partial}

\newcommand{\bea}{\begin{eqnarray}}
\newcommand{\eea}{\end{eqnarray}}
\newcommand{\be}{\begin{equation}}
\newcommand{\ee}{\end{equation}}
\newcommand{\ba}{\begin{align}}
\newcommand{\ea}{\end{align}}

\def\Re{{\rm Re}}
\def\Im{{\rm Im}}

\def\log{{\rm log}}

\def\O{\mathcal{O}}

\def\I{\mathcal{I}}

\def\bh{{\bar h}}
\def\bz{{\bar z}}
\def\bw{{\bar w}}

\def\ba{{\bar a}}

\def\tA{\widetilde A}

\def\th{\widetilde h}

\def\tT{\widetilde T}
\def\tC{\widetilde C}
\def\tN{\widetilde N}
\def\tQ{\widetilde Q}
\def\tDelta{\widetilde \Delta}

\def\th{\widetilde h}

\def\tDelta{\widetilde \Delta}

\def\tq{\widetilde q}

\def\bdelta{\boldsymbol{\delta}}

\def\scri{\mathcal I}

\numberwithin{equation}{section} 

\begin{document}

 \begin{titlepage}
  \thispagestyle{empty}
  \begin{flushright}
  CPHT-RR022.042020
  \end{flushright}
  \bigskip
  \begin{center}
	 \vskip2cm
  \baselineskip=13pt {\LARGE \scshape{Asymptotic Symmetries and Celestial CFT}}
   \vskip1cm 
   \centerline{ 
   {Laura Donnay}${}^\spadesuit {}^\diamondsuit{}^\odot$, 
   {Sabrina Pasterski}${}^\bigstar{}^\diamondsuit$,
   {and Andrea Puhm}${}^\blacklozenge{}^\diamondsuit{}^\odot$
   }

	   \bigskip

  \end{center}

\begin{abstract}
  \noindent

We provide a unified treatment of conformally soft Goldstone modes which arise when spin-one or spin-two conformal primary wavefunctions become pure gauge for certain integer values of the conformal dimension $\Delta$. This effort lands us at the crossroads of two ongoing debates about what the appropriate conformal basis for celestial CFT is and what the asymptotic symmetry group of Einstein gravity at null infinity should be. Finite energy wavefunctions are captured by the principal continuous series $\Delta\in 1+i\mathbb{R}$ and form a complete basis. We show that conformal primaries with analytically continued conformal dimension can be understood as certain contour integrals on the principal series. This clarifies how conformally soft Goldstone modes fit in but do not augment this basis.
Conformally soft gravitons of dimension two and zero which are related by a shadow transform are shown to generate superrotations and non-meromorphic diffeomorphisms of the celestial sphere which we refer to as shadow superrotations. This dovetails the Virasoro and Diff(S$^2$) asymptotic symmetry proposals and puts on equal footing the discussion of their associated soft charges, which correspond to the stress tensor and its shadow in the two-dimensional celestial CFT.

\end{abstract}

\vfill

\noindent{\em${}^\spadesuit$ Institute for Theoretical Physics, Vienna University of Technology,}
{\em A-1040 Vienna, Austria}

\noindent{\em ${}^\diamondsuit$ Center for the Fundamental Laws of Nature, Harvard University,}
{\em Cambridge, MA 02138, USA}

\noindent{\em${}^\odot$ Black Hole Initiative, Harvard University,}
{\em Cambridge, MA 02138, USA}

\noindent{\em ${}^\bigstar$ Princeton Center for Theoretical Science,}
{\em Princeton, NJ 08544, USA}

\noindent{\em${}^\blacklozenge$ CPHT, CNRS, Ecole polytechnique, IP Paris,}
{\em F-91128 Palaiseau, France}

 \end{titlepage}
\tableofcontents

\pagebreak

\section{Introduction}

The realization that soft theorems in gauge theory and gravity are manifestations of Ward identities for asymptotic symmetries~\cite{Strominger:2013lka,Strominger:2013jfa,Strominger:2017zoo} has reinvigorated recent attempts at flat space holography~\cite{deBoer:2003vf} as well as reopened some questions regarding what the physical asymptotic symmetry group of asymptotically flat spacetime should be~\cite{Bondi:1962px,Sachs:1962wk,Sachs:1962zza}. One of the key milestones from applying this paradigm is that it solidifies~\cite{Kapec:2014opa} a conjectured extension~\cite{Banks:2003vp} of the BMS group to include superrotations~\cite{Barnich:2009se,Barnich:2011ct,Barnich:2010eb}, based on a newly discovered subleading soft graviton theorem~\cite{Cachazo:2014fwa}. This precipitated two parallel initiatives within the field.  

On the one hand, the subleading soft graviton mode is a natural stress tensor candidate~\cite{Kapec:2016jld} for a putative dual celestial CFT with a Virasoro symmetry. This provoked reexamining scattering amplitudes in a basis~\cite{Pasterski:2016qvg,Pasterski:2017kqt,Pasterski:2017ylz} that makes conformal covariance manifest which, in turn, demanded a new understanding of the conformal analog of soft theorems~\cite{Cheung:2016iub, Donnay:2018neh,Fan:2019emx,Pate:2019mfs,Adamo:2019ipt,Puhm:2019zbl,Guevara:2019ypd,Law:2019glh,Fotopoulos:2019vac,Banerjee:2020kaa}. While the leading conformally soft graviton naturally arises as the zero mode in this discussion, a puzzling question arose on how the subleading conformally soft graviton would fit into the appropriate conformal basis for celestial amplitudes.

On the other hand, while the leading soft graviton theorem is equivalent to the Ward identity of BMS supertranslation symmetry, the correspondence between the subleading soft graviton theorem and the proposed asymptotic Virasoro superrotation\footnote{By superrotations we collectively mean the angle-dependent generalization of rotations and boosts. They are distinguished from general diffeomorphisms of the celestial sphere.} symmetry does not appear to be bijective. Questions were raised as to whether the asymptotic symmetry group should be extended from the double copy Virasoro symmetry  generated by local conformal Killing vectors (CKVs) to arbitrary diffeomorphisms of the celestial sphere (Diff($S^2$))~\cite{Campiglia:2015yka,Compere:2018ylh,Campiglia:2020qvc}. Unlike for local CKVs, the Ward identity for Diff($S^2$) symmetry appears to be equivalent to the subleading soft graviton theorem~\cite{Campiglia:2014yka}, but the price to pay for this enlarged asymptotic symmetry group is a relaxed set of boundary conditions which modify the metric of the celestial sphere at leading order. 

This paper provides a unified treatment of conformally soft Goldstone modes which lands us at the crossroads of these two ongoing themes and offers a resolution to some of the tensions surrounding what the appropriate conformal basis for celestial amplitudes is, and what the appropriate asymptotic symmetry group for Einstein gravity at null infinity should be. 

Four-dimensional scattering amplitudes exhibit conformal properties when the standard plane wave modes are replaced by so-called conformal primary wavefunctions.  These are labeled by the conformal dimension $\Delta$ and spin $J$ of the representation of the 4D Lorentz group $SL(2,\mathbb{C})$ which acts as the 2D conformal group on the celestial sphere at null infinity.  In~\cite{Pasterski:2017kqt} it was shown that finite energy modes are captured by conformal primary wavefunctions on the principal continuous series $\Delta\in 1+i\mathbb{R}$ and that these form a conformal basis. 
Whereas, in the momentum basis, the leading and subleading soft poles of scattering amplitudes arise in a series expansion near $\omega\rightarrow0$, in the conformal basis, the leading and subleading conformally soft theorems arise from different limits of the conformal dimension $\Delta$~\cite{Cheung:2016iub,Pasterski:2017kqt,Donnay:2018neh,Fan:2019emx,Pate:2019mfs,Adamo:2019ipt,Puhm:2019zbl,Guevara:2019ypd}. 
As shown in~\cite{Donnay:2018neh}, the asymptotic large $U(1)$ Kac-Moody symmetry of gauge theory and the BMS supertranslation symmetry of gravity are generated by respectively spin-one and spin-two conformal primaries with conformal dimension $\Delta=1$. Insertions of these $\Delta=1$ modes into celestial amplitudes give rise to conformally soft factorization theorems in gauge theory~\cite{Fan:2019emx,Nandan:2019jas,Pate:2019lpp} and gravity~\cite{Adamo:2019ipt,Puhm:2019zbl,Guevara:2019ypd}. 

Furthermore, a subleading conformally soft theorem in gravity is obtained~\cite{Adamo:2019ipt,Guevara:2019ypd} in the $\Delta=0$ limit of celestial graviton amplitudes. This corresponds to the insertion of a conformally soft graviton whose conformal dimension does not lie on the principal continuous series. This theme appears to continue for the subsubleading conformally soft theorem in gravity which arises in the $\Delta=-1$ limit of celestial graviton amplitudes~\cite{Guevara:2019ypd}, as well as for the subleading conformally soft theorem for celestial gluon amplitudes which is obtained in the $\Delta=0$ limit~\cite{Adamo:2019ipt}. 
The nature of this analytic continuation in conformal dimension raised some questions. In particular, it was unclear how these modes would fit with the results of~\cite{Pasterski:2017kqt} that the principal continuous series wavefunctions form a complete basis. Here we show how conformally soft primaries with analytically continued conformal dimensions can be understood as certain contour integrals along the principal continuous series. These modes therefore do not augment the conformal basis, even with zero energy modes taken into account.

Related to the subleading $\Delta=0$ conformally soft graviton $h^0$ by a shadow transform is the \mbox{$\tDelta=2-\Delta=2$} spin-two conformal primary $\th^2$. The modes ${h}^{0}$ and $\tilde{h}^{2}$ are both pure diffeomorphisms. Near null infinity, $\tilde{h}^{2}$ matches the form of a superrotation~\cite{Donnay:2018neh} thus supporting an asymptotic Virasoro symmetry. Its relation to the stress tensor for a putative celestial CFT, which is the soft charge, was examined in~\cite{Donnay:2018neh, Ball:2019atb}. On the other hand, ${h}^{0}$ does not obey the standard fall-off conditions. Instead, we will show that it generates an asymptotic Diff$(S^2)$ symmetry. The corresponding soft charge is divergent and a regularization method needs to be employed. Following~\cite{Compere:2018ylh} we take care of radial divergences by adding appropriate boundary terms and demand consistency with the subleading soft graviton theorem.
The renormalized soft Diff$(S^2)$ charge we eventually arrive at turns out to be the shadow transform of the 2D stress tensor.

Why was such a renormalization procedure not needed in computing the soft superrotation charge~\cite{Barnich:2011mi,Kapec:2014opa}? In fact, even superrotations modify the round metric of the celestial sphere at isolated points when the generator is not part of the global $SL(2,\mathbb{C})$. Here we show that when keeping track of these contact terms in the diffeomorphism vector field associated to $\th^2$, a regularization method akin to the one used in defining a soft Diff$(S^2)$ charge~\cite{Compere:2018ylh} is required in order to derive the soft superrotation charge. 

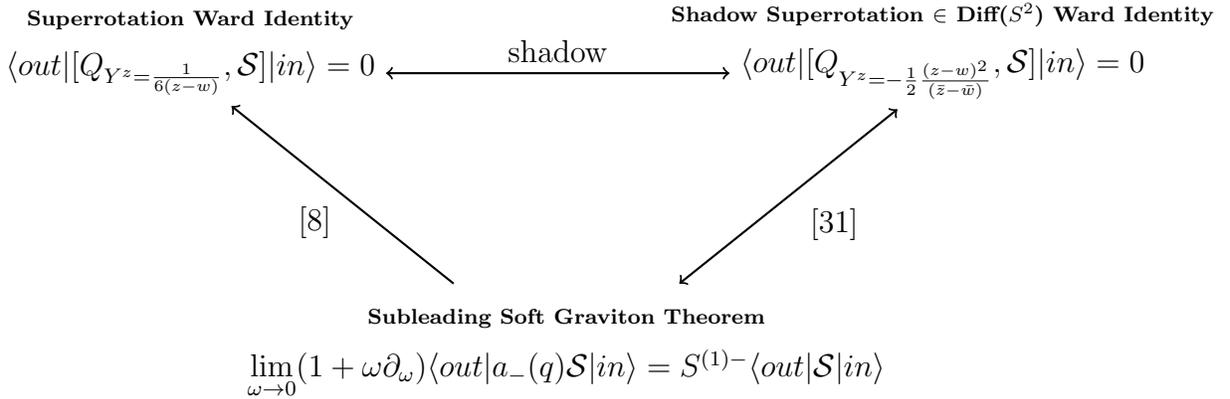
\begin{figure}[h]
\centering ~
\begin{tikzpicture}
\node [label=above:{\bf \scriptsize Superrotation Ward Identity}] (1) at (-5,2) {$\langle out |[Q_{Y^z=\frac{1}{6(z-w)}},\mathcal{S}]|in\rangle=0$  };
\node [label=above:{\bf \scriptsize Shadow Superrotation $\in$ Diff($S^2$) Ward Identity}] (2) at (5,2) {$\langle out |[Q_{Y^z=-\frac{1}{2}\frac{(z-w)^2}{(\bz-\bw)}},\mathcal{S}]|in\rangle=0$  };
\node [label=above:{\bf \scriptsize Subleading Soft Graviton Theorem}] (3) at (0,-2) {$\lim\limits_{\omega\rightarrow 0}(1+\omega\p_\omega)\langle out |a_- (q)\mathcal{S}| in\rangle =S^{(1)-} \langle out| \mathcal{S} | in \rangle$};
\draw[<->,thick] (1)--(2) node[midway,above] {shadow};
\draw[<-,thick] (1)--($(1)!0.7!(3)$) node[midway,below left] {\cite{Kapec:2014opa} };
\draw[<->,thick] (2)--($(2)!0.7!(3)$) node[midway,below right] {\cite{Campiglia:2014yka}};
\end{tikzpicture}
\caption{Superrotations imply the subleading soft graviton theorem so long as the asymptotic Virasoro symmetry group is enhanced to include shadow superrotations.}
\label{fig:ASG}
\end{figure}

This puts the conformally soft gravitons $\th^2$ and $h^0$ on equal footing and dovetails the Virasoro and Diff$(S^2)$ asymptotic symmetry proposals.
In~\cite{Kapec:2014opa} the subleading soft graviton theorem was shown to imply a Ward identity for superrotations within semiclassical $S$-matrix elements which is given by the conformally soft graviton $\th^2$. Campiglia and Laddha then demonstrated in~\cite{Campiglia:2014yka} an equivalence between the subleading soft graviton theorem and the Ward identity for certain Diff($S^2$) transformations which turn out to be none other than the conformally soft graviton ${h}^{0}$. Here we see, that the shadow operation implies that the converse of the relation found in~\cite{Kapec:2014opa} also holds. This is illustrated in Figure~\ref{fig:ASG}. The asymptotic symmetry group for Einstein gravity at null infinity should thus include the closure of Virasoro under shadows within Diff$(S^2)$.  The Ward identity for this enlarged symmetry group is equivalent to the Cachazo-Strominger soft theorem~\cite{Cachazo:2014fwa}. If one allows more general convolutions than the shadow transformation, the asymptotic symmetry group extends to all of Diff$(S^2)$.

We thus come away with a better understanding of both the debate around Virasoro, generated by local CKVs, versus Diff$(S^2)$, generated by arbitrary vector fields, and the interpretation of conformal primaries with analytically continued conformal dimension with respect to the principal continuous series basis. Moreover, the machinery we develop is primed to tackle interpretations of further subleading soft modes~\cite{Adamo:2019ipt,Guevara:2019ypd} which are not pure gauge, as well as the corresponding memory effects in gauge theory and gravity which involves a careful handling of an admixture of incoming and outgoing modes as explained in~\cite{Donnay:2018neh}. We leave these interesting topics to future work.

This paper is organized as follows. We begin by establishing our coordinate conventions in section~\ref{sec:2}.  In section~\ref{sec:cpw}, we review properties of the Mellin transform  (section~\ref{sec:mellin})  and the relevant conformal primary wavefunctions from~\cite{Pasterski:2017kqt}  (sections~\ref{sec:PWtoCP}-\ref{subsec:ConformalBasis}). We build most of the tools we will need in section~\ref{sec:GeneralCPs}.  In section~\ref{subsec:gendelta} we use properties of the Mellin inversion theorem to define a useful distribution that helps us analytically continue results previously established for wavefunctions on the principal series. We discuss practical regularizations of this distribution and then show in section~\ref{subsec:OffPS} how applying them offers a quaint interpretation of modes off the principal series in terms of superpositions of modes on the principal series. This straightens out previous attempts~\cite{Ball:2019atb} at interpreting symplectic pairings between wavefunctions with arbitrary conformal dimensions (section~\ref{subsec:GeneralPairing}) and clarifies the principal series mode expansion~\cite{Pasterski:2017kqt,Donnay:2018neh} which we write down in section~\ref{subsec:ModeExpansion}. We conclude this section by combining these results and tools to define analytically continued mode operators in section~\ref{subsec:Charges} which we connect to soft charges for asymptotic symmetries in section~\ref{sec:CSMem}. Sections~\ref{subsec:SoftPhoton} and~\ref{subsec:SoftGraviton} are mainly a review of the results of~\cite{Donnay:2018neh} which identified the conformal primary modes corresponding to the leading soft photon and soft graviton theorems. In section~\ref{subsec:SubSoftGraviton} we apply the methods of section~\ref{sec:GeneralCPs} to the subleading soft graviton and identify the spin-two Goldstone modes $\th^2$ and $h^0$ with generators of the proposed Virasoro and Diff($S^2$) symmetry groups of Einstein gravity at null infinity. We show how a careful treatment of contact terms puts them on equal footing in computing the soft charges. As a further payoff, this illuminates an under-appreciated connection to the work of~\cite{Campiglia:2014yka,Compere:2018ylh} even for superrotations.

\section{Coordinate Conventions}\label{sec:2}
We begin by outlining our coordinate conventions in this section.  Cartesian coordinates on four-dimensional Minkowski spacetime\footnote{We use signature convention $(-,+,+,+).$} $\mathbb R^{1,3}$ are related to Bondi coordinates $(u,r,z,\bz)$ by the transformation
\begin{equation}
 X^0=u+r\,, \quad X^1=r\frac{z+\bz}{1+z\bz}\,, \quad X^2=- i r \frac{z-\bz}{1+z\bz}\,, \quad X^3=r\frac{1-z\bz}{1+z\bz}\,,
\end{equation}
which maps the Minkowski line element to
\begin{equation}\label{gBondi}
ds^2=-du^2-2du dr+2r^2 \gamma_{z \bar z} dz d\bar z \quad \text{with}\quad \gamma_{z \bar z}=\frac{2}{(1+z \bar z)^2}\,.
\end{equation}
One reaches future null infinity $\scri^+$ by holding $(u,z,\bz)$ fixed and going to large $r$, and we refer to the $S^2$ cross section of future null infinity as the celestial sphere.  Corresponding quantities exist in an expansion near past null infinity where $v=u+2r$ is held fixed in place of $u$, with a corresponding past celestial sphere. For any asymptotic symmetry under consideration only an appropriate diagonal subgroup acting simultaneously on future and past null infinity will be a symmetry of the four-dimensional scattering problem~\cite{Strominger:2017zoo}.

A massless particle crosses the celestial sphere at a point $(w,\bw)$ with momentum 
\begin{equation}\label{pmu}
p^\mu=\omega q^\mu(w,\bw)\,,
\end{equation}
with $\omega \geq0$ and $q^\mu(w,\bw)$ a null vector given by
\begin{equation}\label{qmu}
 q^{\mu}(w,\bw)=(1+w\bw,w+\bw,-i(w-\bw),1-w\bw)\,.
\end{equation} 
Under an $SL(2,\mathbb C)$ transformation 
\begin{equation}
w \to \frac{a w+b}{cw+d} \,, \quad  \bar w \to \frac{ \bar a  \bar w+ \bar b}{ \bar c \bar w+ \bar d}\,,
\label{SL2C}
\end{equation} 
$q^\mu$ transforms as a vector up to a conformal weight\footnote{The energy $p^0=\omega (1+w \bw)$ transforms as $p^0\to p^0\Lambda^0_0$ while $\omega \to \omega|cw+d|^2$.}
\begin{equation}
 q^{\mu}\to  q^{\mu'} =|cw+d|^{-2}\Lambda^\mu_{\,\,\, \nu}q^\nu\,.
\end{equation} 
Here $ad-bc=1$  and $\Lambda^\mu_{\,\;\nu}$ is the associated $SL(2,\mathbb{C})$ group element in the four-dimensional representation. 
Note that null vectors \eqref{qmu} satisfy
\begin{equation}\label{eq:q1}
 q^{\mu}(w,\bw) q_{\mu}(w',\bw')=-2|w-w'|^2\,,
\end{equation} 
and the derivative of~\eqref{qmu} with respect to $w$ ($\bw$) gives the photon polarization vector $\epsilon^\mu_w$ ($\epsilon^\mu_\bw$) of positive (negative) helicity:
\begin{equation} \label{eq:q2}
\partial_w q^\mu=\sqrt 2 \epsilon^\mu_w(q)=(\bw,1,-i,-\bw) \quad \,, \quad \partial_\bw q^\mu=\sqrt 2 \epsilon^\mu_\bw(q)=(w,1,i,-w)\,.
\end{equation}
These satisfy
\begin{equation}\label{eq:q3}
 \epsilon_w\cdot q=0\,, \quad \epsilon_w\cdot \epsilon_w=0\,, \quad \epsilon_w\cdot \epsilon_\bw=1\,,
\end{equation}
and similarly for $w\leftrightarrow \bw$.
The graviton polarization tensor of positive (negative) helicity is $\epsilon^{\mu\nu}_{w}=\epsilon^\mu_w \epsilon^\nu_w$ ($\epsilon^{\mu\nu}_{\bw}=\epsilon^\mu_\bw \epsilon^\nu_\bw$).

\section{Conformal Primary Wavefunctions}\label{sec:cpw}

Four-dimensional scattering amplitudes were recently shown~\cite{Pasterski:2016qvg,Pasterski:2017ylz} to exhibit conformal properties when the standard plane wave basis is replaced by a basis of so-called conformal primary wavefunctions~\cite{Pasterski:2017kqt}. For massless fields this can be achieved via a Mellin transform. In this section we review the construction in~\cite{Pasterski:2017kqt}\footnote{This builds on earlier work~\cite{Costa:2011mg,Costa:2011dw,SimmonsDuffin:2012uy,Costa:2014kfa} using the embedding space formalism.} of conformal primaries in gauge theory and gravity which form a complete conformal basis for single particle states when the spectrum is the principal continuous series of the Lorentz group. 

\subsection{The Mellin Transform}\label{sec:mellin}

The Mellin transform is defined as follows
\begin{equation}\label{Mellin}
\mathcal{M}[f](\Delta)=\int_0^\infty d\omega \omega^{\Delta-1}f(\omega)\equiv \varphi(\Delta)\,,
\end{equation}
with the inverse transform given by 
\begin{equation}\label{InverseMellin}
\mathcal{M}^{-1}[\varphi](\omega)=\frac{1}{2\pi i}\int_{c-i\infty}^{c+i\infty} d\Delta \, \omega^{-\Delta}\varphi(\Delta)=f(\omega)\,.
\end{equation}
The Mellin transform is well defined for functions $f(\omega)$ such that 
\begin{equation}
\int_0^\infty d\omega \omega^{k-1}|f(\omega)|<\infty\,,
\end{equation}
for some $k>0$, and the existence of the inverse transform requires $c>k$.
If these conditions are satisfied then we have the identity
\begin{equation}
\varphi(\Delta)=\mathcal{M}[\mathcal{M}^{-1}[\varphi]](\Delta)\,.
\end{equation}

\subsection{Conformal Primaries from Mellin Transforms of Plane Waves}\label{sec:PWtoCP}
The Mellin transform~\eqref{Mellin} and its inverse transform~\eqref{InverseMellin} allow us to go back and forth between the plane wave basis of the standard momentum-space formulation of massless scattering processes and wavefunctions that transform as conformal primaries under the Lorentz group. 
The relevant pair of functions is given by
\begin{equation}\label{fphi}
\badat{3}
f(\omega)&=e^{\pm i \omega q\cdot X - \varepsilon \omega q^0} \quad \text{with}\quad \varepsilon>0\,,\\
\quad\varphi(\Delta)
&=\frac{\Gamma(\Delta)}{(\pm i)^\Delta} \frac{1}{(-q\cdot X \mp i \varepsilon q^0)^\Delta} \quad \text{with} \quad \mathrm{Re}[\Delta]>0\\
&\equiv \phi^{\Delta,\pm}(X^\mu;w;\bw)\,,
\eadat
\end{equation}
where in the last line we introduced the massless spin-zero conformal primary wavefunction. Spin-one and spin-two massless conformal primary wavefunctions are gauge equivalent to Mellin transforms of plane waves multiplied by the appropriate polarization vectors and tensors as we will discuss in the following.
Note that the $i \varepsilon$-prescription in~\eqref{fphi} is added to make the Mellin integral convergent or to circumvent the singularity at $q\cdot X=0$. This can be achieved by the imaginary timelike shift $X^\mu\to X^\mu_\pm=X^\mu \pm i \varepsilon V^\mu$ with $V^\mu=(-1,0,0,0)$. The appearance of the conformal factor $q^0=1+w \bw$ is a consequence of the parametrization of the momentum~\eqref{pmu}-\eqref{qmu}. In Bondi coordinates $X_\pm^2$ is then regulated at the light-sheet by $u\to u\mp i \varepsilon$.

\subsubsection{Gauge Theory}
The outgoing $(+)$ and incoming $(-)$ massless spin-one conformal primary wavefunctions are~\cite{Cheung:2016iub,Pasterski:2017kqt}
\begin{equation}\label{ADelta}\begin{aligned}
 A^{\Delta,\pm}_{\mu;a}(X^\mu;w,\bw) &= \frac{ (-q\cdot X_\pm)\partial_{a} q_{\mu}+(\partial_{a} q\cdot X_\pm)q_{\mu}}{(-q \cdot X_\pm)^{\Delta+1}}\\
 &=\frac{\Delta-1}{\Delta}\frac{\sqrt 2 (\pm i)^\Delta}{\Gamma(\Delta)}V^{\Delta,\pm}_{\mu;a}+\nabla_\mu \alpha^{\Delta,\pm}_a\,,
\end{aligned}\end{equation}
where $a=w$ or $\bw$ and $V^{\Delta,\pm}_{\mu;a}$ is the Mellin transform of a plane wave
\begin{equation}\label{eq:V1}
 V^{\Delta,\pm}_{\mu;a} =\epsilon_{\mu;a} \phi^{\Delta,\pm}\,,
\end{equation}
with the polarization vector $\epsilon_{\mu;a}=\frac{1}{\sqrt{2}}\p_a q_\mu$ and the residual gauge parameter is
\begin{equation}
\alpha^{\Delta,\pm}_a=\frac{\p_a q\cdot X_\pm}{\Delta (-q\cdot X_\pm)^\Delta}.
\end{equation}
The wavefunctions~\eqref{ADelta} satisfy both Lorenz and radial gauge conditions 
\begin{equation}\label{spin1gaugecond}
 \nabla^\mu  A_{\mu;a}^{\Delta,\pm}=0\,,\quad X^\mu_{\pm} A_{\mu;a}^{\Delta,\pm}=0\,,
\end{equation}
and are solutions to the source-free four-dimensional Maxwell equations, which reduce in this gauge to
\begin{equation}\label{Maxwell}
\nabla_\rho \nabla^\rho A_{\mu;a}^{\Delta,\pm}=0\,,
\end{equation}
with field strength $F_{\mu \nu;a}^{\Delta,\pm}=\nabla_\mu A_{\nu;a}^{\Delta,\pm}-\nabla_\nu A_{\mu;a}^{\Delta,\pm}$ given by
\begin{equation}\label{FDelta}
\badat{2}
F_{\mu \nu;a}^{\Delta,\pm}=\frac{(\Delta-1)(q_\mu\partial_a q_\nu-q_\nu \p_a q_\mu)}{(-q \cdot X_\pm)^{\Delta+1}}\,.
\eadat
\end{equation}

The wavefunctions~\eqref{ADelta} transform as two-dimensional conformal primaries with conformal dimensions $(h,\bar h)=\frac{1}{2}(\Delta+J,\Delta-J)$ 
under an $SL(2,\mathbb{C})$ Lorentz transformation:
\begin{equation}
A_{\mu;a}^{\Delta,\pm}\left(\Lambda^\rho_{\,\, \nu}X^\nu;\frac{a w+b}{c w+d},\frac{\bar a \bw+\bar b}{\bar c \bw+\bar d}\right)=(c w+ d)^{2h}(\bar c \bw+ \bar d)^{2\bh}\Lambda_\mu^{\,\; \sigma}A_{\sigma;a}^{\Delta,\pm}(X^\rho;w,\bw)\,,
\label{SL2Cspin1}
\end{equation}
where the index  $a=w$ corresponds to the spin $J=+1$ (positive helicity) while $a=\bw$ corresponds to $J=-1$ (negative helicity). In two-dimensional conformal field theory, the shadow transform maps a primary with conformal dimension~$\Delta$ to a primary with conformal dimension $\tDelta=2-\Delta$. 
The shadow transform of the spin-one conformal primary wavefunction~\eqref{ADelta} is~\cite{Pasterski:2017kqt}
\begin{equation} \label{SHADelta}
\widetilde{A^{\Delta,\pm}_{\mu;\bar{a}}}=(-X_\pm^2)^{1-\Delta} A_{\mu;a}^{2-\Delta,\pm}\equiv \widetilde{A}^{\tDelta,\pm}_{\mu;a}\,.
\end{equation}

\subsubsection{Gravity}
The outgoing $(+)$ and incoming $(-)$ massless spin-two conformal primary wavefunctions are
\begin{equation}\label{hDelta}\begin{aligned}
 h^{\Delta,\pm}_{\mu\nu;a}(X^\mu;w,\bw)&=\frac{1}{2}\frac{[(-q\cdot X_\pm)\partial_{a} q_{\mu}+(\partial_{a} q\cdot X_\pm)q_{\mu}][(-q\cdot X_\pm)\partial_{a} q_{\nu}+(\partial_{a} q\cdot X_\pm)q_{\nu}]}{(-q\cdot X_\pm)^{\Delta+2}}\, \\
 &=\frac{\Delta-1}{\Delta+1}\frac{(\pm i)^\Delta}{\Gamma(\Delta)} V^{\Delta,\pm}_{\mu\nu;a}+\nabla_\mu \zeta^{\Delta,\pm}_{\nu;a}+\nabla_\nu \zeta^{\Delta,\pm}_{\mu;a}\,,
 \end{aligned}
\end{equation}
where $V^{\Delta,\pm}_{\mu\nu;a}$ is the Mellin transform of a plane wave
\begin{equation}\label{eq:V2}
 V^{\Delta,\pm}_{\mu\nu;a} = \epsilon_{\mu\nu;a} \phi^{\Delta,\pm}\,,
\end{equation}
with polarization tensor $\epsilon_{\mu\nu;a}=\frac{1}{2}\partial_a q_\mu \partial_a q_\nu$,
and the residual diffeomorphism is 
\begin{equation}
\zeta^{\Delta,\pm}_{\mu;a}=\frac{1}{2(\Delta+1)}\left(\frac{\p_a q_\mu (\p_a q\cdot X_\pm)}{(-q\cdot X_\pm)^\Delta}+\frac{1}{2}\frac{q_\mu (\p_a q\cdot X_\pm)^2}{(-q\cdot X_\pm)^{\Delta+1}}\right)\,.
\end{equation} 
They satisfy harmonic and radial gauge conditions 
\begin{equation}\label{spin2gaugecond}
\eta^{\mu \nu}h_{\mu \nu;a}^{\Delta,\pm}=0 \,, \quad \nabla^\mu h_{\mu \nu;a}^{\Delta,\pm}=0 \,, \quad X^\mu_\pm h_{\mu \nu;a}^{\Delta,\pm}=0\,,
\end{equation}
and are solutions to the vacuum linearized Einstein equations, which reduce to
\begin{equation}\label{Einstein}
\nabla_\rho \nabla^\rho  h_{\mu \nu;a}^{\Delta,\pm}(X^\mu;w,\bw)=0\,.
\end{equation}
The wavefunctions~\eqref{hDelta} transform as both a four-dimensional traceless symmetric rank-two tensor and as two-dimensional spin-two conformal primaries with conformal dimension $(h,\bar h)=\frac{1}{2}(\Delta+J,\Delta-J)$ under an $SL(2,\mathbb{C})$ Lorentz transformation:
\begin{equation}
h_{\mu \nu;a}^{\Delta,\pm}\left(\Lambda^\mu_{\,\, \nu}X^\nu;\frac{a w+b}{c w+d},\frac{\bar a \bw+\bar b}{\bar c \bw+\bar d}\right)=(c w+d)^{\Delta+J}(\bar c \bw +\bar d)^{\Delta-J}\Lambda_{\mu}^{\,\, \rho}\Lambda_{\nu}^{\,\, \sigma}h_{\rho 
\sigma;a}^{\Delta,\pm}(X^\mu;w,\bw)\,.
\label{SL2Cspin2}
\end{equation}
The two-dimensional index $a=ww$ corresponds to spin $J=+2$ (positive helicity) while $a=\bw\bw$ corresponds to $J=-2$ (negative helicity).
The shadow transform of the spin-two conformal primary wavefunction~\eqref{hDelta} is \cite{Pasterski:2017kqt}
\begin{equation} \label{SHhDelta}
\widetilde{h^{\Delta,\pm}_{\mu\nu;\bar{a}}}=(-X_\pm^2)^{1-\Delta} h_{\mu\nu;a}^{2-\Delta,\pm}\equiv \widetilde{h}^{\tDelta,\pm}_{\mu\nu;a}\,,
\end{equation}
and has conformal dimension $\tDelta=2-\Delta$. 

\subsection{Conformal Basis}\label{subsec:ConformalBasis}

To determine the values of the conformal dimension $\Delta$ of the spin-one and spin-two conformal primary wavefunctions relevant to physical scattering processes we now turn to a discussion of the conformal basis.
A natural inner product between complex spin-one wavefunctions is
 \begin{equation}\label{IPspin1}
 (A, A')_\Sigma =- i \int d\Sigma^\rho \,\left[ A^{\nu} {F'}_{\rho\nu}^{*} 
 -{A'}^{* \nu} F_{\rho\nu} 
 \right] \,,
 \end{equation}
and between complex spin-two wavefunction is (see e.g.~\cite{Ashtekar:1987tt,Crnkovic:1986ex,Lee:1990nz,Wald:1999wa}) 
\begin{equation}\label{IPspin2}
( h,h')_\Sigma=-i\int d\Sigma^\rho \Big[ h^{\mu \nu} \nabla_\rho h'^{*}_{\,\,\mu \nu}-2h^{\mu \nu} \nabla_\mu h'^{*}_{\,\,\rho\nu}+h\nabla^\mu h'^{*}_{\,\,\rho\mu}-h \nabla_\rho h'^{*}+h_{\rho\mu} \nabla^\mu h'^{*} - (h \leftrightarrow h'^{*})\Big]\,,
\end{equation}
where $\Sigma$ is a Cauchy surface and $h=h^\sigma_{\,\, \sigma}$ vanishes for conformal primaries due to the tracelessness condition~\eqref{spin2gaugecond}.
The inner products for Mellin representatives  $V^{\Delta,\pm}_{\mu;a}$ and $V^{\Delta,\pm}_{\mu\nu;a}$, defined in equations~(\ref{eq:V1}) and~(\ref{eq:V2}) for $\Delta$ on the principal continuous series of the Lorentz group, i.e. $\Delta =1+i\lambda$ with $\lambda \in \mathbb{R}$, were evaluated by one of us and Shao in~\cite{Pasterski:2017kqt}. The result for both spin-one and spin-two on a constant $X^0$ slice is 
\begin{equation}\label{IPspinV}
 (V^{1+i\lambda,\pm}_{a},V^{1+i\lambda',\pm}_{a'})_{\Sigma_0}=\pm (2\pi)^4 \delta_{aa'} \delta^{(2)}(w-w')\delta(\lambda-\lambda')\,,
\end{equation}
and hence pairs $\Delta=1+i\lambda$ modes with their $\Delta=1-i\lambda$ partners.
It was further shown that these form a complete basis for finite energy wavefunctions.\footnote{These inner products are invariant under gauge transformations that vanish sufficiently rapidly.  They are not invariant under large gauge transformations, which are zero energy modes.  This is important when comparing~(\ref{IPspinV}) to~(\ref{IPspin1Sigma}) and (\ref{IPspin2Sigma}).}

\section{General Conformal Dimension}\label{sec:GeneralCPs}

In this section we show that conformal primaries with  $\Delta\in \mathbb{C}$ can be expressed as a superposition of conformal primaries on the principal continuous series $\Delta \in 1+i\mathbb{R}$.   It is then natural to expand general fields and define asymptotic charges purely in terms of primaries on the principal continuous series. We compute the inner product for conformal primaries with general conformal dimension and use it to define operators that shift the gauge field and the metric at null infinity.  We will be particularly interested in special values of $\Delta \in \mathbb{Z}$ corresponding to Goldstone modes of spontaneously broken asymptotic symmetries~\cite{Cheung:2016iub, Pasterski:2017kqt}. These operators will be related to soft asymptotic charges in section~\ref{sec:CSMem}.

\subsection{Generalized Delta Function}\label{subsec:gendelta}
Given the existence of the Mellin transform and its inverse as stated in section~\ref{sec:mellin}, we begin with the identity
\begin{equation}
\varphi(\Delta)=\mathcal{M}[\mathcal{M}^{-1}[\varphi]](\Delta)\,,
\end{equation}
which, using the definitions~\eqref{Mellin}-\eqref{InverseMellin}, is explicitly
\begin{equation}
\badat{2}
\varphi(\Delta)&=\int_0^\infty d\omega \omega^{\Delta-1}\frac{1}{2\pi i}\int_{c-i\infty}^{c+i\infty} dz \,\omega^{-z}\varphi(z)\\
&=\int_{c-i\infty}^{c+i\infty} (-idz)\left(\frac{1}{2\pi}\int_0^\infty d\omega \,\omega^{\Delta-z-1}\right) \varphi(z)\,.
\eadat
\end{equation} 
This leads us to define the distribution
\begin{equation}\label{BoldDelta}
\bdelta(i(\Delta-z))\equiv \frac{1}{2\pi}\int_0^\infty d\omega \omega^{\Delta-z-1}\,,
\end{equation} 
which will serve the role of a generalization of the Dirac delta function to the complex plane with the property
\begin{equation}\label{eq:contour}
\varphi(\Delta)=\int_{c-i\infty}^{c+i\infty} (-idz)\bdelta(i(\Delta-z)) \varphi(z)\,,
\end{equation} 
for $\Delta\in\mathbb{C}$, and $c$ determined by the condition for existence of the Mellin inverse described above.

In what follows, we will need to distinguish the three different conditions the  $z$-contour $c+i\mathbb{R}$  in~\eqref{eq:contour} can obey. \begin{itemize}
\item For {$\bm{ c = \Re(\Delta)}$} we have $\Re(\Delta-z)=0$ in the integrand of \eqref{eq:contour}. In this case, the generalized distribution~\eqref{BoldDelta} reduces to the Dirac delta function with argument $\Im(\Delta-z)$, as can be shown by a change of variables $\omega=e^{x}$. 
\end{itemize}
To discuss the other types of contours, it will be useful to introduce the following regulated versions of the generalized distribution~\eqref{BoldDelta} which converge on the respective contour. 
\begin{itemize}
\item  For {$\bm{ c < \Re(\Delta)}$} we define 
\begin{equation}\label{eq:d1}
\badat{2}
 \bdelta_{\nu,>}(i(\Delta-z))&\equiv \frac{1}{2\pi}\int_0^\infty d\omega \omega^{\Delta-z-1}e^{-\nu \omega}\,, \quad \mathrm{Re}(\Delta-z)>0\\
 &=\frac{1}{2\pi}\nu^{z-\Delta}\Gamma(\Delta-z)\,,
\eadat
\end{equation}
where $\nu>0$.
\item  For {$\bm{ c > \Re(\Delta)}$} we define
\begin{equation}\label{eq:d2}
\badat{2}
 \bdelta_{\nu,<}(i(\Delta-z))&\equiv \frac{1}{2\pi}\int_0^\infty d\omega \omega^{\Delta-z-1}e^{-\nu/\omega}\,, \quad \mathrm{Re}(\Delta-z)<0\\
 &=\frac{1}{2\pi}\nu^{\Delta-z}\Gamma(z-\Delta)\,,
\eadat
\end{equation}
where $\nu>0$.
\end{itemize}
The Gamma functions appearing in these expressions can be analytically continued to the entire complex plane except for an infinite set of poles at the non-positive integers. These expressions are to be understood inside contour integrals of the form~\eqref{eq:contour} which can be computed using Cauchy's theorem from the residues of the Gamma function
\begin{equation}\label{eq:gamres}
  {\rm Res}_{x=-n}\Gamma(x)=\frac{(-1)^n}{\Gamma(n+1)}\,.
\end{equation}

We will now show that the analog of~\eqref{eq:contour} holds when one substitutes~\eqref{eq:d1} or~\eqref{eq:d2} for~\eqref{BoldDelta} and takes the $\nu\rightarrow0$ limit, namely that
\begin{equation}\label{eq:contournu}
\varphi(\Delta)=\lim_{\nu\to 0}\int_{c-i\infty}^{c+i\infty} (-idz)\bdelta_{\nu,\gtrless}(i(\Delta-z)) \varphi(z) \,.
\end{equation}
Let's start with the regularized distribution~\eqref{eq:d1} in~\eqref{eq:contournu}.
First, note that $\Gamma(\Delta -z)\to 0$ as ${\rm Re}(z)\to +\infty$.  We would thus like to close the contour in~\eqref{eq:contournu} to the right.  We will be allowed to do so, so long as $\varphi(z)$ does not grow fast enough to negate or outpace the suppression of the Gamma function along this arc.  We will allow $\varphi(z)$ to have poles at some set of points $\{z_k\}$ in this region.  Since $\Re(\Delta)>c$ we also pick up the residues of the poles of the Gamma function at $z=\Delta+n$ with $n \in \{0,1,2,\dots\}$.  Then, using \eqref{eq:gamres} we arrive at
\begin{equation}\begin{aligned}\label{eq:delworks}
\lim_{\nu\to 0}\int_{c-i\infty}^{c+i\infty} (-idz)\bdelta_{\nu,>}(i(\Delta-z)) \varphi(z) &= \lim_{\nu\to 0} \Big[\sum_{n=0}^\infty \nu^{n} \frac{(-1)^n}{\Gamma(n+1)} \varphi(\Delta+n)\\
&~~~~+ \sum_{k} \nu^{z_k-\Delta}\Gamma(\Delta-z_k)\mathrm{Res}_{z=z_k}\varphi(z) \Big]\\
&=\varphi(\Delta)\,,
\end{aligned}\end{equation}
 where the only contribution to the sum comes from the $n=0$ pole of the Gamma function, so long as $\Re(z_k-\Delta)>0$.\footnote{Note that if there is a $z_k$ such that $z_k=\Delta+n$ for $n\in\mathbb{Z}$ then this corresponds to a double pole in $\Gamma(\Delta-z)\varphi(z)$ rather than two single poles.  This will modify the residue formula but so long as conditions 1. and 2. hold they are damped in the $\nu\rightarrow0$ limit.}  We have shown that~\eqref{eq:contournu} holds for~\eqref{eq:d1} if
 \begin{enumerate}
 \item $\Gamma(\Delta-z)\varphi(z)\rightarrow0$ as $|z|\rightarrow\infty,~\Re(z)>c$,
 \item  $\varphi(z)$ has no poles in the strip $\Re(z)\in[c,\Re(\Delta)]$.
 \end{enumerate}
A similar sequence of steps will reach an analogous conclusion for~\eqref{eq:d2} in~\eqref{eq:contournu}. Closing the contour to the right since $\Gamma(z-\Delta)\to 0$ as $\Re(z)\to -\infty$ we find that~\eqref{eq:contournu} holds for~\eqref{eq:d2} if
 \begin{enumerate}
 \item $\Gamma(z-\Delta)\varphi(z)\rightarrow0$ as $|z|\rightarrow\infty,~\Re(z)<c$,
 \item  $\varphi(z)$ has no poles in the strip $\Re(z)\in[\Re(\Delta),c]$.
 \end{enumerate}
 
We conclude this section with a word of caution about evaluating the regulated distributions~\eqref{eq:d1} or~\eqref{eq:d2} at $\Re(\Delta-z)=0$ as $\nu\rightarrow0$ and expecting to get a Dirac delta function in $\Im(\Delta-z)$. Letting $\Delta-z=iy$ and $k=-\log \nu$,~\eqref{eq:d1} and~\eqref{eq:d2} become
\begin{equation}
\bdelta_{\nu,\gtrless}(i(\Delta-z))=\frac{1}{2\pi}e^{\pm iky}\Gamma(\pm iy)\,,
\end{equation}
and $\lim\nu\rightarrow0$ corresponds to $\lim k\rightarrow\infty$.  This behaves as a distribution in $y$ as follows. 
Over the (infinitesimal as $k\rightarrow\infty$) window $[y-\frac{\pi}{2k},y+\frac{\pi}{2 k}]$ the phase $e^{\pm iky}$ switches sign. So long as $\Gamma(\pm iy)\varphi(\Delta\mp iy)$ is slowly varying in $y$ the contribution of this window to~\eqref{eq:contour} vanishes. Meanwhile, for generic $\varphi$ this condition precisely breaks down near $y=0$ where the Gamma function has a pole and
\begin{equation}\label{yto0limit}
\bdelta_{\nu,\gtrless}(i(\Delta-z))\sim \pm \frac{1}{2\pi iy}(\cos ky \pm i\sin ky)\,.
\end{equation}
The $k\rightarrow \infty$ limit of the imaginary part of this expression, while divergent as $y\rightarrow0$, is still odd in $y$ and its contribution to~\eqref{eq:contour}  will also average to zero over a small window centered at $y=0$. The real part of~\eqref{yto0limit} is a familiar representation of the Dirac delta function with non-trivial support at $y=0$ but multiplied by $\frac{1}{2}$. How did we get $\frac{1}{2}\delta(y)$? What we have really been doing is taking the $c\rightarrow \Re(\Delta)$ limit of~\eqref{eq:d1} on the $c<\Re(\Delta)$ contour (or of ~\eqref{eq:d2} on the $c>\Re(\Delta)$ contour). In doing so the $n=0$ pole of the Gamma function now lies on the integration contour. To insure~\eqref{eq:contournu} still holds, for~\eqref{eq:d1} the $c\rightarrow \Re(\Delta)^-$ contour would need to be deformed by a small arc leaving the pole at $z=\Delta$ to the right of the contour, while for~\eqref{eq:d2} the $c\rightarrow \Re(\Delta)^+$ contour would need to be deformed by a small arc leaving the pole at $z=\Delta$ to the left.

As a final remark, it is worth noting that while the choice of regularizations in~\eqref{eq:d1} and~\eqref{eq:d2} may seem somewhat arbitrary, they actually naturally appear as the $i\varepsilon$ regularization of spacetime singularities in our wavefunctions~\eqref{fphi}. Indeed, in the inner product computation in appendix~\ref{app:IP} we actually see the integrand in~\eqref{eq:d1}, with $\nu$ replaced by $2\varepsilon q^0$, appearing in~\eqref{eq:regdel} before taking the $\varepsilon\rightarrow0$ limit. We can thus expect any analysis involving admixtures of in and out states -- as necessary when examining electromagnetic and gravitational memory in the conformal basis~\cite{Donnay:2018neh,upcoming3} -- to take these regularizations as physical.

\subsection{Conformal Primaries with Analytically Continued $\Delta$}\label{subsec:OffPS} 

Building on the results of the previous section we now show that conformal primary wavefunctions with general conformal dimensions can be expressed as a superposition of wavefunctions on the principal continuous series of the Lorentz group. This result follows after realizing that the conformal primaries~\eqref{ADelta} and~\eqref{hDelta} satisfy the conditions 1. and 2. on $\varphi(z)$ discussed in section~\eqref{subsec:gendelta} for the $c\gtrless \Re(\Delta)$ contours with $c=1$. We can thus substitute $A^{z,\pm}_{\mu;a}$ and $h^{z,\pm}_{\mu;a}$ for $\varphi(z)$ in~\eqref{eq:contournu} and deform the principal series contour.\footnote{This is somewhat reminiscent of the contour deformations employed in the CFT literature to translate between conformal partial waves on the principal series and conformal blocks. While the integration kernel in~\eqref{eq:contourRight} and~\eqref{eq:contourLeft} may be non-standard, its asymptotic behavior gives the desired convergence of the deformed principal series contour for which we recover conformally soft Goldstone modes with integer conformal dimension. Moreover, the regulators in~\eqref{eq:d1} and~\eqref{eq:d2} are natural given the comments at the end of section~\ref{subsec:gendelta}.}

Conformal primaries with with conformal dimension lying to the right of the principal series ${\rm Re}(\Delta)>1$ can be expressed as
\begin{equation}\label{eq:contourRight}
\badat{2}
A_{\mu;a}^{\Delta,\pm}(X^\mu;w,\bw)&= \lim\limits_{\nu\rightarrow0}\frac{1}{2\pi i} \int_{1-i\infty}^{1+i\infty} dz \, \nu^{z-\Delta}\Gamma(\Delta-z)A_{\mu;a}^{z,\pm}(X^\mu;w,\bw)\,,\\
h_{\mu\nu;a}^{\Delta,\pm}(X^\mu;w,\bw)&= \lim\limits_{\nu\rightarrow0}\frac{1}{2\pi i} \int_{1-i\infty}^{1+i\infty} dz \,\nu^{z-\Delta}\Gamma(\Delta-z)h_{\mu\nu;a}^{z,\pm}(X^\mu;w,\bw)\,.
\eadat
\end{equation}
To verify this, note that the $z$-dependence of the integrand for either spin takes the form
\begin{equation}
\left(\frac{\nu}{-q\cdot X_\pm}\right)^z\Gamma(\Delta-z)\rightarrow 0\,, \quad \mathrm{Re}(z)\rightarrow+\infty\,.
\end{equation}
Applying the results of section~\ref{subsec:gendelta} we close the contour to the right of the principal continuous series, picking up the residues of the poles of the Gamma function at $z=\Delta+n$ for $n=\{0,1,2,\dots\}$ while there are no poles in our corresponding $\varphi(z)$ to worry about. Using
\begin{equation}
\sum_{n=0}^\infty y^{\Delta+n}\frac{(-1)^n}{\Gamma(n+1)}=e^{-y}y^\Delta,
\end{equation}
we arrive at 
\begin{equation}
 \lim\limits_{\nu\rightarrow 0} \nu^{-\Delta}\exp\left[-\left(\frac{\nu}{-q\cdot X_\pm }\right)\right]\left(\frac{\nu}{-q\cdot X_\pm }\right)^\Delta =\frac{1}{(-q\cdot X_\pm)^\Delta}\,,
 \end{equation}
thus validating~\eqref{eq:contourRight}. An analogous line of arguments holds for the shadow primaries~\eqref{SHADelta} and~\eqref{SHhDelta} for $\Re(\tDelta)>1$ with the same contour.
 
\begin{figure}[ht]
\begin{subfigure}{.5\textwidth}
  \centering
  \includegraphics[width=.7\linewidth]{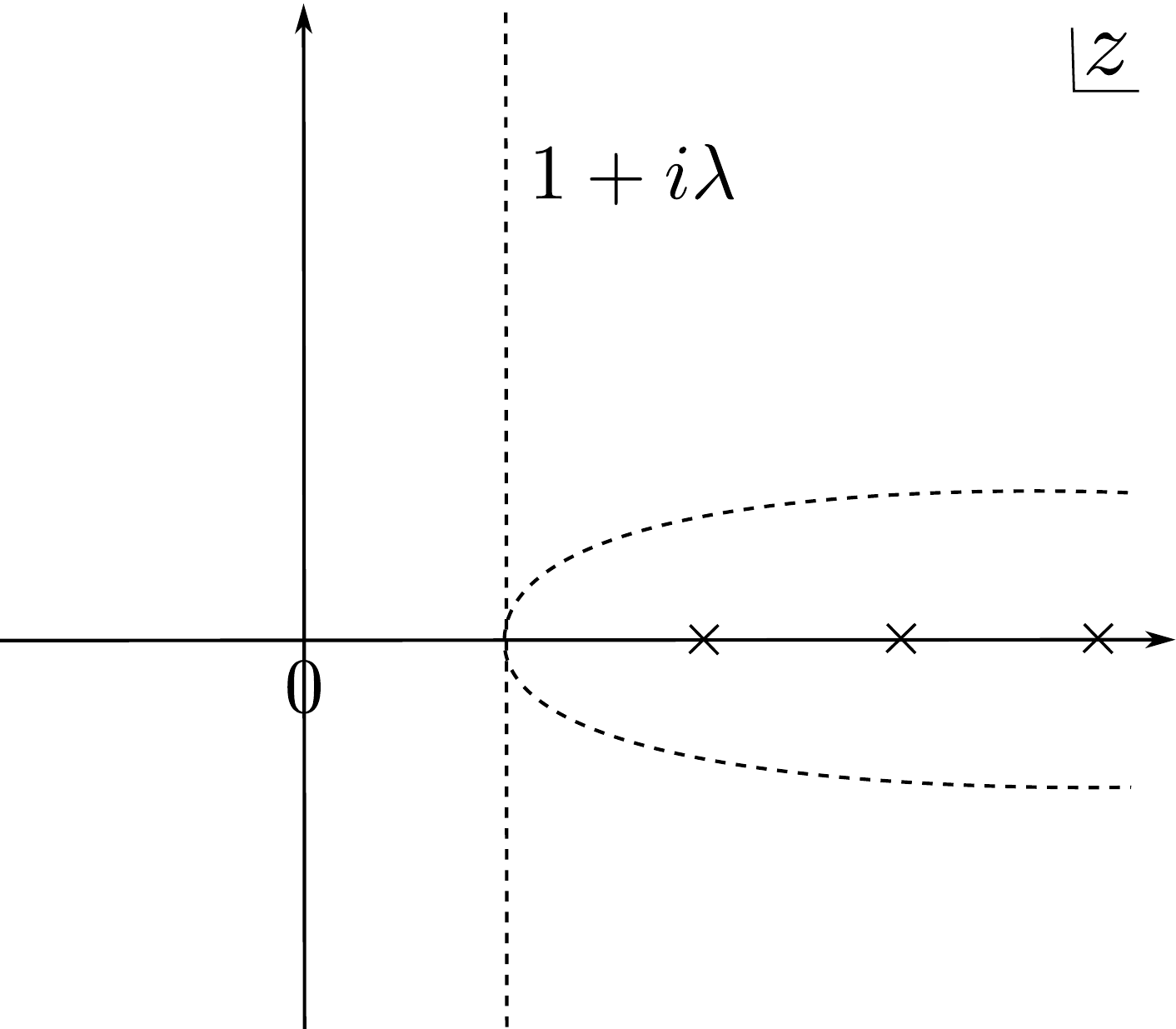}  
  \caption{Contour deformation for ${\rm Re}(\Delta)>1$.}
  \label{fig:right}
\end{subfigure}
\begin{subfigure}{.5\textwidth}
  \centering
  \includegraphics[width=.7\linewidth]{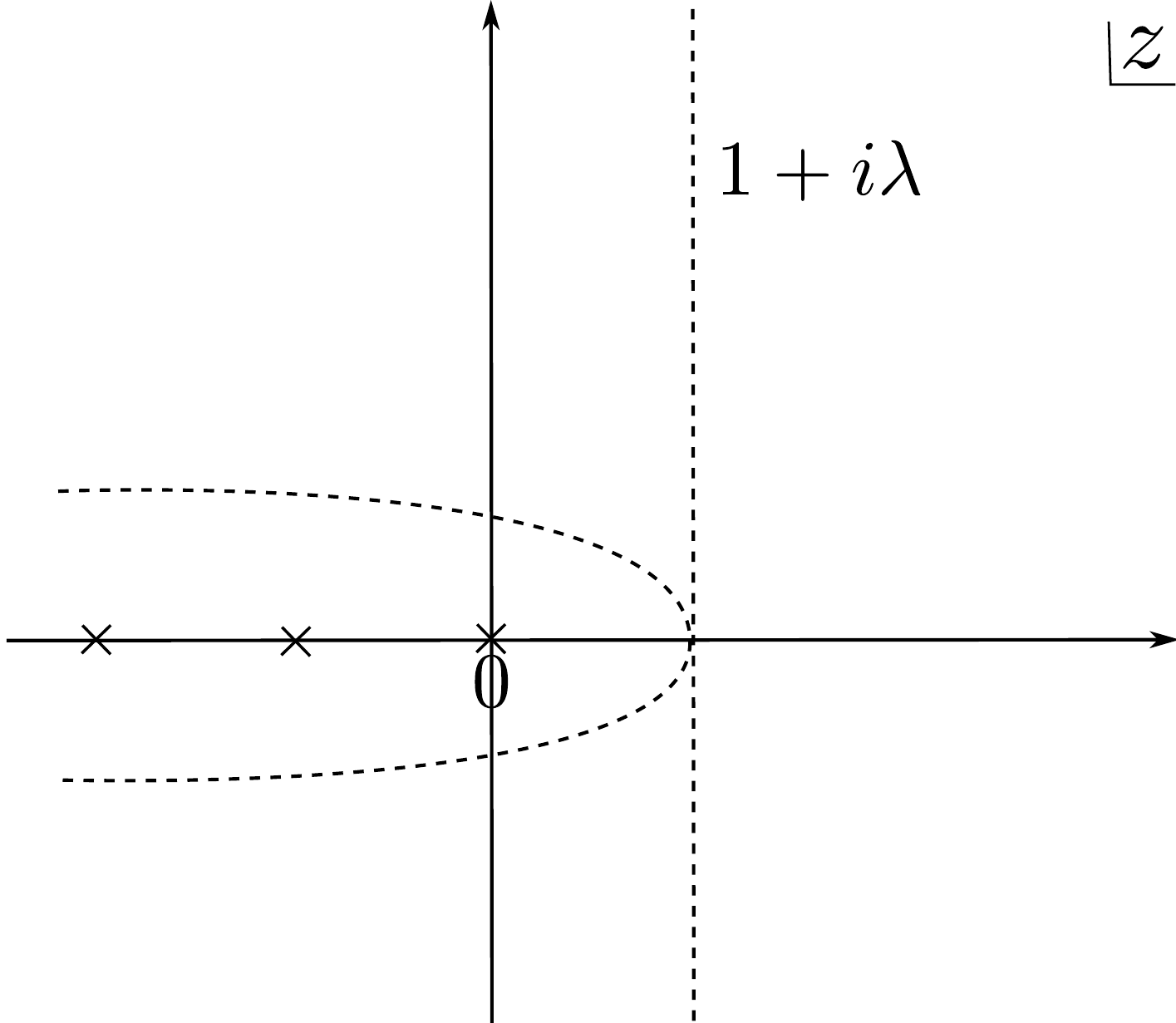}  
  \caption{Contour deformation for ${\rm Re}(\Delta)<1$.}
  \label{fig:left}
\end{subfigure}
\caption{Conformal primaries with general conformal dimension $\Delta$ can be expressed as integrals along the principal series contour $z=1+i\lambda$ with $\lambda \in \mathbb{R}$. Depending on whether ${\rm Re}(\Delta) \gtrless 1$ the contour is deformed to the right (\ref{fig:right}) or left (\ref{fig:left}).  For illustration we have drawn the pole positions corresponding to $\Delta=2$ and $\Delta=0$ in (\ref{fig:right}) and (\ref{fig:left}), respectively. Using~\eqref{SHADelta} and~\eqref{SHhDelta}, we see the same contours are relevant for the shadow modes with $\tilde{\Delta}=2$ and $\tilde{\Delta}=0$, respectively.}
\label{fig:fig}
\end{figure}

Conformal primary wavefunctions with conformal dimension lying to the left of the principal series ${\rm Re}(\Delta)<1$ can be expressed as
\begin{equation}\label{eq:contourLeft}
\badat{2}
A_{\mu;a}^{\Delta,\pm}(X^\mu;w,\bw)&= \lim\limits_{\nu\rightarrow0}\frac{1}{2\pi i} \int_{1-i\infty}^{1+i\infty} dz \, \nu^{\Delta-z}\Gamma(z-\Delta)A_{\mu;a}^{z,\pm}(X^\mu;w,\bw)\,,\\
h_{\mu\nu;a}^{\Delta,\pm}(X^\mu;w,\bw)&= \lim\limits_{\nu\rightarrow0}\frac{1}{2\pi i} \int_{1-i\infty}^{1+i\infty} dz \, \nu^{\Delta-z}\Gamma(z-\Delta)h_{\mu\nu;a}^{z,\pm}(X^\mu;w,\bw)\,.
\eadat
\end{equation}
In this case,
 \begin{equation}
\left(\nu(-q\cdot X\mp i\varepsilon)\right)^{-z}\Gamma(z-\Delta)\rightarrow 0\,, \quad\mathrm{Re}(z)\rightarrow-\infty\,,
\end{equation}
and we thus close the contour to the left picking up the residues of the poles of the Gamma function at $z=\Delta-n$ for $n=\{0,1,2,\dots\}$ while there are again no poles in the corresponding $\varphi(z)$. We get a sum of the form
\begin{equation}
\sum_{n=0}^\infty y^{-\Delta+n}\frac{(-1)^n}{\Gamma(n+1)}=e^{-y}y^{-\Delta}\,,
\end{equation}
and so 
\begin{equation}
 \lim\limits_{\nu\rightarrow 0} \nu^{\Delta}\exp\left[-\nu(-q\cdot X_\pm)\right]\left(\nu(-q\cdot X_\pm )\right)^{-\Delta} =\frac{1}{(-q\cdot X_\pm)^\Delta}\,,
 \end{equation}
validating~\eqref{eq:contourLeft}. The same arguments can be repeated for the shadow primaries~\eqref{SHADelta} and~\eqref{SHhDelta} for $\Re(\tDelta)<1$.

We have thus shown that conformal primary wavefunctions~\eqref{ADelta} and~\eqref{hDelta} (and their shadows~\eqref{SHADelta} and~\eqref{SHhDelta}) with arbitrary conformal dimension can be expanded in terms of conformal primaries on the principal series.

\subsection{Inner Product}\label{subsec:GeneralPairing}

We are now ready to discuss the inner product for conformal primary wavefunctions with arbitrary conformal dimension $\Delta$.  Relegating the explicit computations to Appendix~\ref{app:IP}, we reach the following results.\footnote{\label{ft:pair}Note that the pairings between the $\Delta=1$ Goldstone modes of gauge theory and gravity and their respective canonical partners constructed in~\cite{Donnay:2018neh}, namely the conformally soft photon and graviton, are not captured by the righthand side of~\eqref{IPspin1Sigma} and~\eqref{IPspin2Sigma}. They can be readily obtained though from the combination of incoming and outgoing wavefunctions discussed in~\cite{Donnay:2018neh} using the regulated versions~\eqref{eq:d1}~-~\eqref{eq:d2} of the distribution~\eqref{BoldDelta}.}
The inner product~\eqref{IPspin1} between two spin-one primaries~\eqref{ADelta} is \begin{equation}\label{IPspin1Sigma}
(A_{a}^{\Delta,\pm}(w), A_{a'}^{\Delta',\pm}(w'))_{\Sigma} 
= \pm 2 (2\pi)^4e^{\pm i \pi \Delta}\sin(\Delta\pi) \frac{(\Delta-1)}{\pi\Delta(\Delta-2)}\delta_{aa'}\delta^{(2)}(w-w')\bdelta(i(\Delta+\Delta'^*-2))\,.
\end{equation}
For spin-two primaries defined in~\eqref{hDelta} the inner product~\eqref{IPspin2} is
\begin{equation}\label{IPspin2Sigma}
(h_{a}^{\Delta,\pm}(w),h_{a'}^{\Delta',\pm}(w'))_{\Sigma}
=\pm (2\pi)^4 e^{\pm i \pi \Delta}\sin(\Delta\pi) \frac{(\Delta-1+i)(\Delta-1-i)}{\pi\Delta(\Delta-1)(\Delta-2)}\delta_{aa'}\delta^{(2)}(w-w')\bdelta(i(\Delta+\Delta'^*-2))\,.
\end{equation}
As demonstrated in appendix~\ref{app:IP} these expressions hold  regardless of whether the Cauchy slice $\Sigma$ is a constant $X^0$ slice or taken to null infinity.
The $\sin(\pi\Delta)$ appearing in the inner products introduces zeros for integer $\Delta$ except where cancelling poles appear, i.e. $\Delta=\{0,2\}$ for spin-one, and $\Delta=\{0,1,2\}$ for spin-two. 

While orthogonality holds on the principal continuous series~\cite{Pasterski:2017kqt}, the discussion in section~\ref{subsec:gendelta} and~\ref{subsec:OffPS} highlights that we should not take the appearance of the distribution~\eqref{BoldDelta} in~\eqref{IPspin1Sigma}-\eqref{IPspin2Sigma} as an orthogonality condition that pairs modes of weight $\Delta$ with modes of weight $2-\Delta^*$ for arbitrary $\Delta\in \mathbb{C}$. Indeed, the regulated distributions~\eqref{eq:d1}-\eqref{eq:d2} generically give a non-zero inner product between any two $\{\Delta,\Delta'\}\in\mathbb{C}$. Moreover, it makes sense that the inner product between two such wavefunctions would generically be non-trivial since they can each be viewed as a wavepacket of principal series wavefunctions. One should keep this caveat in mind when interpreting the `pairings' in~\cite{Ball:2019atb} of spin-two modes away from the principal series with dimensions zero and two. 
Besides its relevance for the subleading soft graviton mode which we examine in section~\ref{subsec:SubSoftGraviton}, this observation is also important for further subleading soft modes corresponding to subleading soft photon and sub-subleading soft graviton modes~\cite{Adamo:2019ipt,Guevara:2019ypd} as will be addressed in~\cite{upcoming2}.

\subsection{Mode Expansion}\label{subsec:ModeExpansion}
 
Now that we understand that conformal primary wavefunctions with arbitrary conformal dimensions can be expressed as a superposition of wavefunctions with $\Delta\in1+i\mathbb{R}$, it is natural to expand general fields purely on the principal continuous series of the Lorentz group.
An arbitrary gauge field perturbation is given by\footnote{Subtleties involving zero modes were examined by two of us and Strominger in~\cite{Donnay:2018neh}. Memory effects, discussed there for $\Delta=1$ conformally soft modes, involve admixtures of incoming and outgoing modes. A proper treatment of those in the mode expansions~\eqref{eq:ex11} and~\eqref{eq:ex22} goes beyond the scope of this paper; we will return to it in~\cite{upcoming3}.} 
\begin{equation}\label{eq:ex11}
\badat{2}
A_{\mu}(X)&= \int d^2 w \sum_{b=w,\bw}\int_{1-i\infty}^{1+i\infty} (-id\Delta)\, \Big[\mathcal{N}^+_{1,2-\Delta}A_{\mu;b^*}^{2-\Delta,+}(X^\mu;w,\bw) a_{1,b}(\Delta;w,\bw)\\
&\qquad \qquad \qquad \qquad \qquad \qquad+\mathcal{N}^-_{1,{\Delta}} A_{\mu;b}^{\Delta,-}(X^\mu;w,\bw) a_{1,b} (\Delta;w,\bw)^\dagger\Big]\,,
\eadat
\end{equation}
and an arbitrary metric perturbation is given by
\begin{equation}\label{eq:ex22}
\badat{2}
h_{\mu\nu}(X)&=\int d^2 w \sum_{b=w,\bw} \int_{1-i\infty}^{1+i\infty}(-i d\Delta) \,\Big[\mathcal{N}^+_{2,2-\Delta} h_{\mu\nu;b^*}^{2-\Delta,+}(X^\mu;w,\bw)  a_{2,b}(\Delta;w,\bw)\\
&\qquad \qquad \qquad \qquad \qquad \qquad +\mathcal{N}^-_{2,\Delta} h_{\mu\nu;b}^{\Delta,-}(X^\mu;w,\bw)  a_{2,b}(\Delta;w,\bw)^\dagger \Big]\,.
\eadat
\end{equation}
Here we have introduced the normalization factors
\begin{equation}
\mathcal{N}^\pm_{1,\Delta}=\left[2 (2\pi)^4 e^{\pm i \pi \Delta}\sin(\Delta \pi) \frac{(\Delta-1)}{\pi \Delta(\Delta-2)}\right]^{-1/2}\,,
\end{equation}
and
\begin{equation}
\mathcal{N}^\pm_{2,\Delta}=\left[(2\pi)^4 e^{\pm i \pi \Delta}\sin(\Delta\pi) \frac{(\Delta-1+i)(\Delta-1-i)}{\pi \Delta(\Delta-1)(\Delta-2)}\right]^{-1/2}\,,
\end{equation}
so that the creation and annihilation operators with spin $j=\{1,2\}$ obey the commutation relations 
\begin{equation}
[a_{j,a}(\Delta,w,\bw),a_{j,a'}(\Delta',w',\bw')^\dagger]= \delta_{aa'}\delta^{(2)}(w-w')\bdelta(i(\Delta+\Delta'^*-2))\,,
\end{equation}
where we have used $\mathcal{N}^\pm_{j,2-\Delta}=\mathcal{N}^\mp_{j,\Delta}$. On the principal continuous series, where $\Delta=1+i\lambda$, $\Delta'= 1+i \lambda'$ with $\lambda,\lambda' \in \mathbb{R}$, this reduces to the standard Dirac delta function $\delta(\lambda-\lambda')$.

\subsection{Analytically Continued Quantum Modes}\label{subsec:Charges}

Let us now define the following operators\footnote{Note that conformal primaries with analytically continued conformal dimension away from $\Delta\in 1+i\mathbb{R}$ do not obey the standard fall-off conditions near null infinity. The associated operators~\eqref{Qscri} are divergent and need to be renormalized. In section~\ref{sec:CSMem} we will outline this for the conformally soft gravitons $h^0$ and $\th^2$.}
\begin{equation}\label{Qscri}
Q^{\Delta}_{1,a}(w,\bw)\equiv i(A,A_{a^*}^{\Delta^*,+}(w,\bw))_{\Sigma}\,, \quad Q^{\Delta}_{2,a}(w,\bw)\equiv i(h,h_{a^*}^{\Delta^*,+}(w,\bw))_{\Sigma}\,.
\end{equation}
Using the mode expansions~\eqref{eq:ex11}-\eqref{eq:ex22}, the results for the inner products~\eqref{IPspin1Sigma}-\eqref{IPspin2Sigma}, and the relation~\eqref{eq:contour}, these reduce to 
\begin{equation}\label{Qmode}
Q^{\Delta}_{1,a}=i(\mathcal{N}^-_{1,\Delta})^{-1}a_{1,a}(\Delta;w,\bar{w})\,, \quad 
Q^{\Delta}_{2,a}=i(\mathcal{N}^-_{2,\Delta})^{-1}a_{2,a}(\Delta;w,\bar{w})\,.
\end{equation}
These operators shift\footnote{Here we have used~\eqref{eq:contour}, $\mathcal{N}^\pm_{j,2-\Delta}=\mathcal{N}^\mp_{j,\Delta}$  and the fact that on the principal continuous series $\Delta^*=2-\Delta$.}  the gauge field as
\begin{equation}\label{shift1}
[Q^{\Delta}_{1,a}(w,\bw),A_{\mu}(X)]=iA_{\mu;a}^{\Delta,-}(X^\mu;w,\bw)\,,
\end{equation}
and the metric as
\begin{equation}\label{shift2}
[Q^{\Delta}_{2,a}(w,\bw),h_{\mu\nu}(X)]=ih_{\mu\nu;a}^{\Delta,-}(X^\mu;w,\bw) \,.
\end{equation}
In particular, this implies that acting with these operators on the vacuum produces single particle states with conformal primary wavefunctions of dimension $\Delta$:
\begin{equation}
\langle 0|Q^{\Delta}_{1,a}(w,\bw) A_{\mu}(X)|0\rangle=iA_{\mu;a}^{\Delta,-}(X^\mu;w,\bw),
\end{equation}
and
\begin{equation}
\langle 0|Q^{\Delta}_{2,a}(w,\bw)h_{\mu\nu}(X)|0\rangle=ih_{\mu\nu;a}^{\Delta,-}(X^\mu;w,\bw).
\end{equation}
The notation for the operators $Q_{j,a}^\Delta$ is intended to be suggestive. In the following section, we will show how for certain values of $\Delta$ for which the conformal primary is pure gauge, the operators $Q_{j,a}^\Delta$ correspond to soft charges in the full (matter-coupled) theory when we take the Cauchy slice $\Sigma $ to future null infinity $\mathcal{I}^+$.\footnote{For past null infinity, the soft part of the charge transforming the in state will be related to particular $(Q_{j,a}^\Delta)^\dagger$.}

\section{Asymptotic Symmetries}\label{sec:CSMem}

Soft theorems in gauge theory and gravity imply Ward identities for corresponding asymptotic symmetries~\cite{Strominger:2017zoo}. Whenever there is a gauge transformation that obeys some boundary conditions but acts non-trivially at the boundary, one can use the canonical formalism \cite{Wald:1999wa, Crnkovic:1986ex} to construct a non-zero charge associated to it.  When Stokes' theorem is used to express this charge as an integral along null infinity, the part of the charge linear in the fields is referred to as the soft charge~\cite{Strominger:2017zoo}. While the soft theorem/Ward identity connection becomes nontrivial only in the full (coupled) theory, the inner product of an arbitrary field perturbation with a Goldstone mode is enough to identify the soft part of the charge.  Indeed the soft charge is the operator which generates an inhomogeneous shift of the gauge field tangent to the asymptotic symmetry.

It is thus natural to try to  map any residual gauge transformations to soft charges.   In~\cite{Pasterski:2017kqt} it was shown that for certain values of the conformal dimension the conformal primary wavefunctions $A^\Delta$ and $h^\Delta$, defined in~\eqref{ADelta} and~\eqref{hDelta}, and their respective shadows $\tA^{2-\Delta}$ and $\th^{2-\Delta}$, defined in~\eqref{SHADelta} and~\eqref{SHhDelta}, reduce to pure gauge or diffeomorphism modes. By inspecting their large $r$ behavior near null infinity, these can be seen to correspond to Goldstone modes of spontaneously broken asymptotic symmetries of gauge theory and gravity. This is summarized in table~\ref{table:Goldstone}. 
\begin{table}[h!]
\renewcommand*{\arraystretch}{1.3}
\centering
\begin{tabular}{|c|c|cc|}
\hline
  & \multicolumn{1}{c|}{${A}^{\Delta}_{\mu}$}  & \multicolumn{2}{c|}{${h}^{\Delta}_{\mu\nu}$}\\
  \hline
 $\Delta$ &1  &  1&  0 \\
 symmetry & large $U(1)$  & supertranslation & shadow superrotation $\in$ Diff$(S^2)$\\
\hline
\end{tabular}
\caption{Goldstone modes of spontaneously broken asymptotic symmetries of gauge theory and gravity.  The corresponding shadow modes $\tA^{\tDelta=2-\Delta}_{\mu}$ and $\th^{\tDelta=2-\Delta}_{\mu\nu}$ are also pure gauge. While $A^{\Delta=1} _{\mu }$ and $h^{\Delta=1} _{\mu\nu}$ are their own shadows, the shadow of $h^{\Delta=0}_{\mu\nu}$ is the superrotation mode $\tilde{h}^{\tDelta=2} _{\mu\nu}$.}
 \label{table:Goldstone}
\end{table}

In the following sections we will show that we can identify soft charges in gauge theory and gravity with the operators $Q_{j,a}^\Delta$ defined in section \ref{subsec:Charges},  which by~\eqref{shift1}-\eqref{shift2} generate an inhomogeneous shift in the gauge field or metric by the Goldstone modes of table~\ref{table:Goldstone}.  For the leading soft theorems, the computations in sections~\ref{subsec:SoftPhoton} and~\ref{subsec:SoftGraviton} will essentially be a review of results in~\cite{Donnay:2018neh}, involving the Goldstone modes $A^1_\mu=\tA^1_\mu$ and $h^1_{\mu\nu}=\th^1_{\mu\nu}$, to which we add that now we can interpret these charges in terms of conformal primary creation operators by equating the corresponding charges to $Q_{j,a}^{\Delta}$ for spin $j=1,2$ and $\Delta=1$ and then using~\eqref{Qmode}.

The subleading soft graviton case, which we address in section~\ref{subsec:SubSoftGraviton}, is more subtle. It turns out that there are two Goldstone modes, ${h}^{\Delta=0} _{\mu\nu}$ and $\tilde{h}^{\tDelta=2} _{\mu \nu}$, which are related by a shadow transform. Certain aspects of these modes have already been studied in~\cite{Donnay:2018neh} and~\cite{Ball:2019atb}, but the results of section~\ref{sec:GeneralCPs} add an important insight: the subleading soft graviton can be viewed either as an analytic continuation away from the principal series or as a superposition of radiative (on the principal series) modes. Unlike the leading soft graviton, these analytically continued modes are not each canonically paired with only a single other mode as is clear from~\eqref{IPspin2Sigma} thus generalizing the result of~\cite{Ball:2019atb}. 
Our main focus in section~\ref{subsec:SubSoftGraviton} will be the identification of these two Goldstone modes with the generators of celestial conformal symmetry. A crucial point is that previous treatments of the superrotation vector field associated with the Virasoro asymptotic symmetry group have dropped contact terms which are radially divergent and modify the round metric of the celestial sphere at isolated points. This is reminiscent of the action of general diffeomorphisms on the celestial sphere which have been proposed as a distinct extension of the original BMS group. Here, we will show that the superrotation primary $\th^2$ and the Diff($S^2$) primary $h^0$, which we refer to as a shadow superrotation, appear on equal footing in a careful treatment of contact terms.  We evaluate the operator $Q_{j=2,a}^{\Delta}$ and its shadow for $\tDelta=2$ and $\Delta=0$, interpret them in terms of the soft part of the superrotation and Diff($S^2$) charges~\cite{Barnich:2011mi,Strominger:2017zoo,Campiglia:2014yka,Compere:2018ylh} after a suitable renormalization procedure, and discuss their relation to the 2D stress tensor and its shadow.

\subsection{Large $U(1)$ Kac-Moody Symmetry}\label{subsec:SoftPhoton}

For conformal dimension $\Delta=1=\tDelta$ the spin-one conformal primary~\eqref{ADelta} and its shadow~\eqref{SHADelta} degenerate to the same Goldstone mode
\begin{equation}\label{A1}
 {A}^{1,\pm}_{\mu;a}=\partial_\mu \alpha_a^1=\tA^{1,\pm}_{\mu;a}\,,
\end{equation}
with
\begin{equation}
 \alpha_a^{1,\pm}=-\partial_a {\rm log}(-q\cdot X_{\pm})\,.
\end{equation}
At future null infinity~$\I^+$ the angular components of the Goldstone mode $A^{1,\pm}_{\mu;a}$ for positive helicity ($a=w$) takes the form
\begin{equation}\label{A1zandzb}
 A^{1}_{z;w}=-\frac{1}{(z-w)^2}\,, \quad \quad A^{1}_{\bz;w}=2\pi \delta^{(2)}(z-w)\,,
\end{equation}
where we dropped the $\pm$ label since the there no longer is a branch cut. The temporal and radial components of the gauge field near null infinity behave, respectively, as $A^1_{u;w}\sim \O(1/r)$ and $A^1_{r;w}\sim \O(1/r^2)$. Hence, $A^1_{\mu;w}$ obeys the standard fall-off conditions\footnote{One indeed usually (see e.g.~\cite{Strominger:2017zoo}) takes the fall-offs $A_u\sim \O(1/r),\, A_z\sim \O(1),\, A_\bz\sim \O(1)$. 
}
 and obviously has a vanishing field strength.
We recognize the angular expressions~\eqref{A1zandzb} as the Goldstone modes of spontaneously broken large $U(1)$ gauge symmetry
\begin{equation}\label{A1Goldstone}
 A^{1}_{z;w}=\partial_z \varepsilon_w\,, \quad \quad  A^{1}_{\bz;w}=\partial_\bz \varepsilon_w\,,
\end{equation}
where the large gauge parameter $\varepsilon_w$ is the boundary value of $\alpha^1_w$ on $\scri^+$
\begin{equation}\label{varepsilon}
\varepsilon_w=\frac{1}{z-w}\,.
\end{equation}
Evaluating~\eqref{Qscri} for $\Delta=1$ and $a=w$ we find\footnote{The superscript $(n)$ denotes the order $\frac{1}{r^n}$ at which the fields appear in the large $r$ expansion.} 
\begin{equation}
\badat{2}
Q^1_{1,w}\equiv i(A,A_\bw^{\Delta=1})_{\scri^+}
&=-\int du d^2 z \, \left( A^1_{z;w} F_{u\bz}^{(0)}+A^1_{\bz;w} F_{uz}^{(0)} \right).
\eadat
\end{equation}
Comparing this with the definition for the soft charge for the corresponding asymptotic $U(1)$ Kac-Moody symmetry \cite{He:2014cra,Strominger:2017zoo}
\begin{equation}\label{qedl}
\badat{2}
Q_{\varepsilon}^{soft}&=-\frac{1}{e^2}\int du  d^2 z \, (\partial_z \varepsilon\, F_{u\bz}^{(0)}  + \partial_\bz \varepsilon \, F_{uz}^{(0)})\,,
\eadat
\end{equation}
we find (after reinstating the factor $1/e^2$ in the definition of the charge operator)
\begin{equation}
\boxed{Q^1_{1,w}=Q_{\varepsilon}^{soft}\textstyle{[\varepsilon_w=\frac{1}{z-w}]}}\,.
\end{equation}

\subsection{BMS Supertranslation Symmetry}\label{subsec:SoftGraviton}

Supertranslations are generated by arbitrary functions on the sphere $f=f(z,\bz)$ and take the form (see e.g. \cite{Strominger:2017zoo})
\begin{equation}\label{STdiffeo}
\zeta_f=f \partial_u + \frac{1}{2} D^2 f \partial_r -\frac{1}{r} D^A f \partial_A +\dots \,,
\end{equation}
where sphere indices $A=(z,\bz)$ are raised and lowered with the round sphere metric $\gamma_{AB}$ and its inverse, while $D_A$ and $D^2=D_A D^A$ denote, respectively, the covariant derivative and the Laplacian with respect to $\gamma_{AB}$. The action of supertranslations \eqref{STdiffeo} on the free gravitational data $C_{AB}$ defined by $g_{AB}=r^2 \gamma_{AB}+r C_{AB}+\dots$ is
\begin{equation}\label{deltaf}
 \delta_f C_{AB}=f\partial_u C_{AB}-2D_A D_B f+\gamma_{AB} D^2 f \,.
\end{equation}
In~\cite{Donnay:2018neh} it was shown that the inhomogeneous part of the transformation~\eqref{deltaf},
\begin{equation}\label{STshift}
 \delta_f^{\rm shift} C_{AB}\equiv -2D_A D_B f+\gamma_{AB} D^2 f\,,
\end{equation}
is generated by spin-two conformal primaries with dimension $\Delta=1=\tDelta$, where the conformal primaries and their shadows again degenerate to the same Goldstone mode 
\begin{equation}\label{h1}
 {h}^{1,\pm}_{\mu\nu;a}=\nabla_\mu \zeta_{\nu;a}^{1,\pm}+\nabla_\nu \zeta_{\mu;a}^{1,\pm}=\th^{1,\pm}_{\mu\nu;a}\,,
\end{equation}
with diffeomorphism vector
\begin{equation}\label{xi1}
\zeta_{\mu;a}^{1,\pm}=-\frac{1}{8} \partial_a^2[q_\mu {\rm log} (-q\cdot X_{\pm})]\,.
\end{equation}

Near future null infinity $\scri^+$, the vector field $\eta^{\mu\nu}\zeta^{1,\pm}_{\nu;a}\partial_\mu$ becomes the generator of supertranslations~\eqref{STdiffeo} with supertranslation parameter for positive helicity $a=ww$ given by
\begin{equation}\label{f}
 f_{ww}=-\frac{1}{4}\frac{(\bz-\bw)}{(z-w)(1+z\bz)}\,.
\end{equation}
The leading components of the metric no longer have a branch cut that we have to regulate and so we will drop the $\pm$ label in the following.
In terms of~\eqref{f}, the gravitational data given by the angular component of the Goldstone mode~\eqref{h1} for positive helicity $a=ww$,
\begin{equation}
 C^{\rm 1}_{zz;ww}=\frac{(\bz-\bw)}{(z-w)^3 (1+z\bz)}\,, \quad \quad C^{\rm 1}_{\bz\bz;ww}= \frac{\pi\delta^{(2)}(z-w)}{(1+z\bz)}\,,
\end{equation} 
can be written as 
\begin{equation}
  C^{\rm 1}_{zz;ww}=-2D_z^2 f_{ww}\,, \quad \quad C^{\rm 1}_{\bz\bz;ww}=-2D_\bz^2 f_{ww}\,.
\end{equation}
This corresponds precisely to a pure shift~\eqref{STshift} induced by large diffeomorphisms on $\scri^+$. Hence~\eqref{h1} is the Goldstone mode of spontaneously broken BMS supertranslation symmetry. The remaining metric components obey the standard fall-off conditions\footnote{Usually, one fixes Bondi gauge and demands the fall-offs $h_{uu}\sim \O(1/r)$, $h_{ur}\sim \O(1/r^2),\, h_{uz}\sim \O(1)$, $h_{u\bz}~\sim~\O(1),\, h_{zz}\sim \O(r), \, h_{z\bz}\sim \O(1),\, h_{\bz\bz}\sim \O(r)$. 
} and the Bondi news tensor defined as $N^1_{AB;ww}=\partial_u C^1_{AB;ww}$ vanishes.

Evaluating $Q^1_{2,ww}$ gives
\begin{equation}
\badat{2}
Q^1_{2,ww}\equiv i(h,h_{\bw \bw}^{\Delta=1})_{\scri^+}&=-\int du d^2 z \sqrt{\gamma}\, \left( C^{1}_{zz;w w}  N^{zz}+C^{1}_{\bz \bz;w w} N^{\bz\bz}\right).
\eadat
\end{equation}
Comparing this with the definition for the soft charge for the corresponding supertranslation symmetry \cite{He:2014laa,Strominger:2017zoo}
\begin{equation}
\badat{2}
Q_{f}^{soft}&=-\frac{1}{16\pi G}\int du  d^2 z \sqrt{\gamma}\, (D_z^2 f N^{zz}+D_{\bz}^2 f N^{\bz\bz}),
\eadat
\end{equation}
we find (after reinstating the factor $1/(16\pi G)$ in the definition of the charge operator)
\begin{equation}
\badat{2}
\boxed{Q^1_{2,ww}=Q_{f}^{soft}\textstyle{[f_{ww}=-\frac{1}{4}\frac{(z-w)}{(\bz-\bw)(1+z\bz)}]}}\,.
\eadat
\end{equation}

\subsection{Superrotation and Diff($S^2$) Symmetry}\label{subsec:SubSoftGraviton}
Besides the $\Delta=1$ Goldstone modes for gauge theory and gravity, for spin-two there exist additional Goldstone modes away from the principal series. These are spin-two conformal primaries with conformal dimension $\Delta=0$
\begin{equation}\label{h0}
 {h}^{0,\pm}_{\mu\nu;a}=\nabla_\mu \xi_{\nu;a}^{0,\pm}+\nabla_\nu \xi_{\mu;a}^{0,\pm} \quad \text{with} \quad  \xi^{0,\pm}_{\mu;a}=-\frac{1}{4} (-q \cdot X_{\pm}) \,\p_a [q_\mu \p_a \log(-q\cdot X_{\pm})]\,,
\end{equation}
and shadow dimension $\tDelta=2-\Delta=2$\,,
\begin{equation}\label{th2}
\th^{2,\pm}_{\mu\nu;a}=\nabla_\mu \xi_{\nu;a}^{2,\pm}+\nabla_\nu \xi_{\mu;a}^{2,\pm} \quad \text{with} \quad \xi^{2,\pm}_{\mu;a}=-\frac{1}{24}\p_a^3[X_\pm^\rho(q_\rho \p_{\bar a}q_\mu-q_\mu \p_{\bar a}q_\rho)\log(-q\cdot X_{\pm})]\,,
\end{equation}
which are related by the shadow transform~\eqref{SHhDelta} as
\begin{equation}
\widetilde{h^{0,\pm}_{\mu\nu;\bar{a}}}= (-X_\pm^2) h^{2,\pm}_{\mu\nu;a}\equiv\th^{2,\pm}_{\mu\nu;a}\,.
\end{equation}
The vector fields in \eqref{h0} and \eqref{th2}\footnote{They should not be confused with the vector field $\zeta^\Delta$ defined in~\eqref{hDelta}. E.g. $\zeta^{\Delta=0}$ is related to the vector field in~\eqref{h0} via $\xi^0_{\nu;a}= \zeta^0_{\nu;a}-\frac{1}{4} \partial_a q_\nu (\partial_a q\cdot X)$.} both satisfy the condition $\Box \xi$=0 and are related by~\cite{Cheung:2016iub}
\begin{equation}
\partial_{\bar{a}} \xi^{2,\pm}_{\mu;a}=-\frac{1}{6}\partial_a^3 \xi^{0,\pm}_{\mu;\bar{a}}\,.
\end{equation}
We will show below that these correspond to particular Diff$(S^2)$ vector fields parameterized by the leading data in~\eqref{Y2} and~\eqref{Y0}. 

In harmonic gauge, diffeomorphisms of the celestial sphere are generated by the vector field
\begin{equation}\label{SRdiffeo}
 \xi_Y=u \alpha \partial_u -\left(\alpha r +u\left(\frac{D^2}{2}+2\right)\alpha \right) \partial_r  + \left(Y^A+\frac{u}{2r} ((D^2+1)Y^A-2D^A\alpha)\right) \partial_A +\dots \,,
\end{equation}
where $Y^A=Y^A(z,\bz)$ is an arbitrary vector field on the sphere and we have introduced $\alpha\equiv\frac{1}{2}D_C Y^C$. A detailed derivation of the form \eqref{SRdiffeo} is presented in Appendix~\ref{app:DiffS2} where we discuss various subtleties that arise when $Y^A$ is not a conformal Killing vector on the sphere, as well as the appearance of logarithmic terms at subleading orders, which are necessary for the diffeomorphism~\eqref{SRdiffeo} to satisfy the harmonic gauge condition.\footnote{See references \cite{Avery:2015gxa,Campiglia:2016efb,Himwich:2019qmj,Compere:2019odm} for other analyses of residual gauge transformations in harmonic gauge.}
It is important to notice that Diff$(S^2)$ vector fields are over-leading, in the sense that they modify the metric of the celestial sphere. Indeed, the action of the Lie derivative along \eqref{SRdiffeo} on a metric of the form $g_{AB}=r^2 \gamma_{AB}+r C_{AB}+\dots$ is given by
\begin{equation}\label{dY}
\badat{2}
&\delta_Y \gamma_{AB}=D_A Y_B+D_B Y_A-2\alpha \gamma_{AB}\,,\\
&\delta_Y C_{AB}=(\alpha u\p_u+\mathcal L_Y-\alpha) C_{AB}-2uD_A D_B \alpha+uD_A (D^2+1)Y_B.
\eadat
\end{equation} 
At this point, an important comment is in order. The first proposed extension of the BMS group in the literature~\cite{Barnich:2009se,Barnich:2011ct,Barnich:2010eb,Kapec:2014opa} considered superrotations which are local conformal Killing vectors that enhance the Lorentz group to two copies of the Virasoro algebra. The surface charges associated to superrotations as well as their algebra were derived in \cite{Barnich:2011mi} (see also~\cite{Barnich:2013axa,Barnich:2017ubf}). Because they violate the CKV condition $D_A Y_B+D_B Y_A=2\alpha \gamma_{AB}$ locally, these meromorphic superrotations change the celestial sphere metric by adding singularities at isolated points. This was physically interpreted in~\cite{Strominger:2016wns} as due to cosmic strings piercing null infinity. As we will see in the following, keeping careful track of these singular terms puts Virasoro on equal footing with Diff($S^2$) as far as their Ward identities are concerned. Indeed, the $\Delta=0$ and $\tilde \Delta=2$ Goldstone primaries~\eqref{h0} and~\eqref{th2} which we will in the following relate to Diff($S^2$) and Virasoro symmetry, respectively, are related by a shadow transform, and so are their corresponding soft charges. 

The action on the news tensor $N_{AB}=\p_u C_{AB}$ can be read from~\eqref{dY}:
\begin{equation}
\badat{2}
&\delta_Y N_{AB}=(\alpha u\p_u+\mathcal L_Y) N_{AB}-2D_A D_B \alpha+D_A (D^2+1)Y_B\,.
\eadat
\end{equation} 
The inhomogeneous parts of the transformations above induce the following shift on the metric functions:
\begin{equation}\label{shift}
\badat{2}
&\delta_Y^{\mathrm{shift}} \gamma_{AB}\equiv D_A Y_B+D_B Y_A\,,\\
&\delta_Y^{\mathrm{shift}} C_{AB}\equiv-2uD_A D_B \alpha+uD_A (D^2+1)Y_B\,,\\
&\delta_Y^{\mathrm{shift}} N_{AB}\equiv -2D_A D_B \alpha+D_A (D^2+1)Y_B\,.
\eadat
\end{equation} 
We will now show that the primaries~\eqref{th2} and~\eqref{h0} are the harmonic gauge diffeomorphisms of the celestial sphere which generate pure shift transformations \eqref{shift} for given $Y^A=Y^A(z,\bz)$.
Finally, we will compute the operator~\eqref{Qscri} for both Goldstone modes and identify it with the soft part of the canonical charge for superrotation and Diff($S^2$) symmetry.

\subsubsection{$\tDelta=2$ Goldstone Mode}
We begin by quoting the gravitational data for the $\tDelta=2$ Goldstone mode~\eqref{th2} for positive helicity $a=ww$
\begin{equation}
  \tC^2_{zz;ww}= \frac{u}{(z-w)^4}\,, 
\end{equation}
which implies the Bondi news tensor $ \tN^{2}_{zz;ww}=\partial_u \tC^2_{zz;ww}$,
\begin{equation}
  \tN^2_{zz;ww}= \frac{1}{(z-w)^4}\,.
\end{equation}
The news is conformally soft as defined in~\cite{Donnay:2018neh} as it transforms as a primary with conformal weights $(h,\bh) = (2,0)$ under an $SL(2,\mathbb C)$ transformation. It moreover has the form of a pure (meromorphic) superrotation;
indeed, near future null infinity $\scri^+$, the vector field $\xi^{2\,\mu}_{a} \partial_\mu$ for $a=ww$ matches the expansion of~\eqref{SRdiffeo} with
\begin{equation}\label{Y2}
 Y^{z}_{ww}=\frac{1}{6(z-w)}\,, \quad Y^\bz_{ww}=0\,,
\end{equation}
which is a (complexified) superrotation. As already pointed out in~\cite{Donnay:2018neh}, the Bondi news tensor can be expressed as
\begin{equation}
 \tN^{2}_{zz;ww}=-D_z^3 Y^z_{ww}\,,
\end{equation}
which is nothing but the pure shift transformation in \eqref{shift} with $Y^A$ given by \eqref{Y2}.
The superrotation vector field~\eqref{Y2} violates the CKV condition at isolated points, and thus changes the celestial sphere metric by adding singularities:
\begin{equation}
\delta_Y \gamma_{\bz\bz}=\mathcal L_Y \gamma_{\bz\bz}=2\gamma_{z\bz}D_\bz Y^z_{ww}=\frac{2\pi}{3} \gamma_{z\bz}\delta^{(2)}(z-w).
\end{equation}
This inhomogeneous shift $\delta_Y^{\rm shift}\gamma_{\bz\bz}$ coincides with the leading  $\mathcal{O}(r^2)$ term in the expansion of the primary near $\scri^+$
\begin{equation}\label{tq2}
\th^2_{\bz\bz;ww}=r^2 \tq^2_{\bz\bz;ww}+r \tC^2_{\bz\bz;ww}+\dots\,,
\end{equation}
while the $\mathcal{O}(r)$ term coincides with the inhomogeneous shift $\delta_Y^{\rm shift} C_{\bz\bz}$ in the gravitational data 
\begin{equation}\label{tC}
\tC^2_{\bz\bz;ww}=-uD_\bz^2 D_z Y^z_{ww}+uD_\bz(D^2+1)Y_{\bz;ww}=u\frac{\pi}{3} \gamma_{z\bz} \left(1+\frac{1}{2}\frac{(1+z\bw)^4}{(1+w\bw)^2} \partial_z \partial_\bz \right)\delta^{(2)}(z-w)\,.
\end{equation}
We thus see that the spin-two primary~\eqref{th2} is the Goldstone mode of spontaneously broken superrotation symmetry.\footnote{As in~\cite{Kapec:2016jld} a generic superrotation can be constructed via a contour integral in the reference direction $w$.} 

\subsubsection{$\Delta=0$ Goldstone Mode}
A shadow transformation takes the positive helicity mode $\th^2_{\mu\nu;ww}$ to the negative helicity mode $h^0_{\mu\nu;\bw\bw}$ with gravitational data
\begin{equation}
 C^0_{zz;\bw\bw}=2\pi u \delta^{(2)}(z-w)\,, \quad C^0_{\bz\bz;\bw\bw}=-2 u\frac{(1+w\bz)(1+z\bw)}{(1+z\bz)^2} \frac{(z-w)}{(\bz-\bw)^3}\,,
\end{equation}
and Bondi news tensor $N^0_{zz;\bw\bw}=\partial_u C^0_{zz;\bw\bw}$ and $N^0_{\bz\bz;\bw\bw}=\partial_u C^0_{\bz\bz;\bw\bw}$.
Near future null infinity, the diffeomorphism $\xi^{0\,\mu}_a\partial_\mu$ for $a=\bw\bw$ takes the form~\eqref{SRdiffeo}  
with the vector field
\begin{equation}\label{Y0}
 Y^z_{\bw\bw}=-\frac{1}{2}\frac{(z-w)^2}{(\bz-\bw)}\,,\quad Y^{\bz}_{\bw \bw}=0.
\end{equation}
This diffeomorphism generates the shifts
\begin{equation}\label{N0}
 N^0_{zz;\bw\bw}=-D^3_z Y^z_{\bw\bw}\,,\quad N^0_{\bz\bz;\bw\bw}=\gamma_{z\bz} D_\bz(D^2+1)Y^z_{\bw\bw}-D_\bz^2 D_z Y^z_{\bw\bw}\,,
\end{equation}
at $\O(r)$ while the leading $\mathcal O(r^2)$ term in 
\begin{equation}
h^{0}_{\bz\bz;\bw\bw}=r^2 q^0_{\bz\bz;\bw\bw}+r C^0_{\bz\bz;\bw\bw}+\dots\,,
\end{equation}
is generated by
\begin{equation}
 \mathcal L_Y \gamma_{\bz\bz}=2\gamma_{z\bz} D_\bz Y^z_{\bw\bw}= \gamma_{z\bz}\frac{(z-w)^2}{(\bz-\bw)^2}\,.
\end{equation}
Hence, the spin-two primary \eqref{h0} is the Goldstone mode of spontaneously broken shadow superrotation $\in$ Diff($S^2$) symmetry. Moreover, the vector field~\eqref{Y0} is the one needed in \cite{Campiglia:2014yka} to show that the Cachazo-Strominger subleading soft graviton theorem can be obtained from a Diff$(S^2)$ Ward identity.

\subsubsection{Soft Charges for Superrotation and Diff($S^2$) Symmetries}

The soft part of the charges associated to superrotations and shadow superrotations~$\in$~Diff($S^2$) is obtained from the inner product~\eqref{IPspin2} at null infinity of a generic metric perturbation $h$ with the Goldstone mode~\eqref{th2} or~\eqref{h0}. However, because the Goldstone modes $h^0$ and $\th^2$ violate the standard Bondi fall-offs, the inner product is divergent and we need to employ a suitable renormalization procedure that takes care of radially divergent terms. A remaining ambiguity in defining the soft charge is fixed by demanding consistency with the subleading soft graviton theorem. We outline these steps now.

In the language of the covariant phase space formalism (see~\cite{Compere:2018aar} for a pedagogical review), computing the inner product~\eqref{IPspin2} at future null infinity amounts to computing the Iyer-Wald symplectic structure~\cite{Iyer:1994ys,Lee:1990nz} 
\begin{equation}\label{IWOmega}
\Omega[\delta g, \delta' g;g]=\int_{\scri^+} \omega[\delta g, \delta' g;g]\,,
\end{equation}
for the variations $\delta g=h$, $\delta' g=h'$ around a fixed background $g$ and the presymplectic form is a spacetime co-dimension one form\footnote{With the notation $\omega=\omega^\rho (d^3x)_\rho$ and $(d^3x)_\rho=\frac{1}{3!}  \varepsilon_{\rho \mu \nu \sigma}dx^\mu \wedge dx^\nu \wedge dx^\sigma$.} given by
\begin{equation}\label{omrho}
\omega^\rho[\delta g, \delta' g;g]= \sqrt{-g} (h^{\mu \nu} \nabla^\rho h'_{\,\,\mu \nu}-2h^{\mu \nu} \nabla_\mu h'^\rho_{\,\,\,\nu}- (h \leftrightarrow h'))\,.
\end{equation}
We are interested in computing~\eqref{omrho} with the metric perturbation $h'$ corresponding to either of the Goldstone modes $h^0_{\mu\nu;\bw\bw}$ or $\th^2_{\mu\nu;ww}$.
When computing the presymplectic form (see Appendix~\ref{app:Qspin2} for details), we find that it contains a radially divergent piece proportional to the inhomogeneous shift~\eqref{shift} of the celestial sphere metric. This feature was already noticed by Comp\`ere, Fiorucci and Ruzziconi in the analysis of Diff$(S^2)$ symmetries~\cite{Compere:2018ylh} and can be expected as the latter are over-leading symmetries in that they change the round metric of the celestial sphere.

A renormalization of the symplectic structure is therefore required in order to obtain finite boundary charges. We will follow the procedure developed in~\cite{Compere:2018ylh}, which exploits the fact that the presymplectic form can be shifted by a boundary term, which expresses a residual ambiguity in the definition of the presymplectic structure in the covariant phase space formalism~\cite{Iyer:1994ys}. After renormalization, the symplectic structure is finite and we find (see Appendix~\ref{app:Qspin2})
\begin{equation}
i(h,h'^*)^{ren}_{\scri^+}=\Omega^{ren}[\delta g,\delta' g;g]\equiv-\delta \mathcal Q_Y\,,
\end{equation}
where
\begin{equation}\label{mQY}
\badat{2}
\mathcal Q_Y=\int du d^2 z  \sqrt \gamma \, (-D_z^3 Y^z  C^{zz}-\frac{1}{2}D^2 D_\bz Y_\bz  C^{\bz \bz}+u D_z^3 Y^z  N^{zz}-\frac{u}{2}D^2 D_\bz Y_\bz  N^{\bz \bz})\,.
\eadat
\end{equation}
There subsists a remaining ambiguity in adding boundary terms at null infinity which can be fixed by demanding consistency with the subleading soft graviton theorem. This amounts to adding the boundary term\footnote{Conformal primary wavefunctions obey the harmonic and a further residual radial gauge condition which is incompatible with the Bondi gauge fixing employed in the literature in the discussion of Diff($S^2$) symmetry and soft charges. We have therefore adapted the renormalization procedure of~\cite{Compere:2018ylh} to our gauge fixing.}
\begin{equation}\label{dQ}
\badat{2}
\Delta \mathcal Q_Y=\int d^2z \sqrt \gamma\, u\,(D_A Y_B+D_B D_C D_A Y^C) C^{AB}\,,
\eadat
\end{equation}
to~\eqref{mQY} and defines the soft charge
\begin{equation}\label{QYsoft}
\badat{3}
Q_Y^{soft}=\mathcal Q_Y+ \int du \p_u \Delta \mathcal Q_Y\,.
\eadat
\end{equation}
For $Y^A=(Y^z,0)$, as is the case for both $\th^2_{\mu\nu;ww}$ and $h^0_{\mu\nu;\bw\bw}$, we have
\begin{equation}\label{udQ}
\badat{2}
\int du \p_u \Delta \mathcal Q_Y=\int du d^2z \sqrt \gamma\, (D_z^3 Y^z  C^{zz}+\frac{1}{2}D^2 D_\bz Y_\bz  C^{\bz \bz}+u D_z^3 Y^z  N^{zz}+\frac{u}{2}D^2 D_\bz Y_\bz  N^{\bz \bz})\,.
\eadat
\end{equation}
This yields the final result for the soft charge associated with (shadow) superrotations:
\begin{equation}\label{QYfinal}
\badat{3}
Q_Y^{soft}=\frac{1}{16\pi G}\int du d^2z \sqrt \gamma\,u D_z^3 Y^z  N^{zz}\,,
\eadat
\end{equation}
with $Y^z=Y^z(z,\bz)$ and we reinstated $1/(32\pi G)$ in the standard definition of the soft charge.

A few comments are in order. The expression for the soft charge~\eqref{QYfinal} which we derived in harmonic gauge takes the same form in Bondi gauge~\cite{Campiglia:2015yka,Distler:2018rwu,Compere:2018ylh}. Moreover, the above counterterm procedure highlights the importance of carefully keeping track of contact terms. 
For superrotations generated by the Goldstone mode $\th^2$ with positive helicity ($a=ww$) we have
\begin{equation}\label{Dz3Ysuperrot}
D_z^3 Y^z_{ww}=-\frac{1}{(z-w)^4}\,,
\end{equation} 
while the radially divergent term in~\eqref{IWOmega} as well as the remaining terms in~\eqref{mQY} are contact terms that get subtracted. In hindsight, this explains why previous discussions of the soft superrotation charge got away with dropping contact terms altogether. The situation is inverted for shadow superrotation symmetry where the final expression for the soft charge generated by the Goldstone mode $h^0$ with negative helicity ($a=\bw\bw$) involves a contact term
\begin{equation}\label{Dz3YDiffS2}
D_z^3 Y^z_{\bw\bw}=-2\pi \delta^{(2)}(z-w)\,,
\end{equation} 
while the radially divergent term in~\eqref{IWOmega} and the remaining terms in~\eqref{mQY} are non-meromorphic functions on the celestial sphere. Clearly, the renormalization procedure is needed to arrive at a sensible soft charge. Moreover, the contact term~\eqref{Dz3YDiffS2} is what establishes the equivalence between the Diff($S^2$) Ward identity and the subleading soft graviton theorem~\cite{Campiglia:2014yka}. 

Finally, we turn to the analytically continued mode operators defined in section~\ref{subsec:Charges} and their role in the putative celestial CFT. 
For the superrotation Goldstone mode $\th^2$ the asymptotic charge operator~\eqref{Qscri} for $\tDelta=2-\Delta=2$ and $a=ww$ defined by $\tQ^2_{2;ww}\equiv i(h,\th^2_{\bw \bw})_{\scri^+}$ corresponds, after renormalization~\eqref{dQ} and~\eqref{udQ}, to the superrotation charge~\eqref{QYfinal} 
\begin{equation}
\boxed{\tQ^{2\,ren}_{2,ww}=Q_{Y}^{soft}\textstyle{[Y_{ww}^z=\frac{1}{6(z-w)},Y^\bz_{ww}=0]}}\,.
\end{equation}
This quantity appears as the 2D stress-tensor for 4D gravity~\cite{Kapec:2016jld, Donnay:2018neh}
\begin{equation}\label{Tww}
T_{ww}\equiv \frac{-6 i}{8 \pi G}\int du d^2z \sqrt{\gamma}\widetilde{N}^2_{zz;ww}u N^{zz}= 12i\,\tilde{Q}^{2\, ren}_{2,ww}\,,
\end{equation}
which is a weight $(h,\bh)=(2,0)$ operator in the putative celestial CFT. 

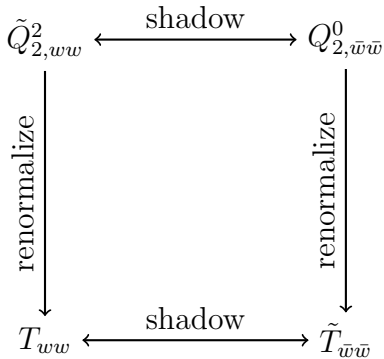
\begin{figure}[h]
\centering 
\begin{tikzpicture}
\node [] (1) at (-2,2) {$\tilde{Q}^{2}_{2,ww}$  };
\node [] (2) at (2,2) {$Q^{0}_{2,\bw\bw}$};
\node [] (3) at (-2,-2) {$T_{ww}$};
\node [] (4) at (2,-2) {$\tilde{T}_{\bw\bw}$};
\draw[<->,thick] (1)--(2) node[midway,above] {shadow};
\draw[<->,thick] (3)--(4) node[midway,above] {shadow};
\draw[->,thick] (1)--(3) node[midway,above,rotate=90] {renormalize};
\draw[->,thick] (2)--(4) node[midway,above,rotate=90] {renormalize};
\end{tikzpicture}
\caption{The shadow~\cite{Dolan:2011dv} and renormalization~\cite{Compere:2018ylh} procedures commute. The renormalized quantities $\{T_{ww},\tT_{\bw\bw}\}$ correspond to soft charges appearing in~\cite{Kapec:2014opa} and~\cite{Campiglia:2014yka}, respectively.}
\label{fig:QshadowT}
\end{figure}

The asymptotic charge operator~\eqref{Qscri} for $\Delta=0$ and $a=\bw\bw$, defined by $Q^0_{2;\bw\bw}\equiv i(h,h^0_{ww})_{\scri^+}$, corresponds after renormalization  to the soft Diff($S^2$) charge
\begin{equation}
\boxed{Q^{0\,ren}_{2,\bw\bw}=  Q_{Y}^{soft}{\textstyle[Y^z_{\bw\bw}=-\frac{1}{2}\frac{(z-w)^2}{(\bz-\bw)}, Y^\bz_{\bw\bw}=0]}}\,.
\end{equation}
Defining the weight $(h,\bh)=(-1,1)$ shadow stress tensor
\begin{equation}
\badat{2}
\tT_{\bw\bw}&\equiv \frac{3}{2\pi}\int d^2 w' \frac{(w-w')^2}{(\bw-\bw')^2}T_{w'w'}\,,
\eadat
\end{equation}
we find\footnote{Here we used that the positive helicity Goldstone mode $\th^2_{zz;ww}$ is related to the negative helicity Goldstone mode $h^0_{zz;\bw\bw}$ by a shadow transform on the celestial sphere where orders of $r$ are not mixed, and made use of the result of~\cite{Dolan:2011dv} that
\begin{equation}\label{I2}
 \badat{2}
\int d^2 z \prod_{i=1}^2 \frac{1}{(z-z_i)^{h_i}}\frac{1}{(\bz-\bz_i)^{\bar h_i}}=\frac{\Gamma(1-h_1)\Gamma(1-h_2)}{\Gamma(\bar h_1)\Gamma(\bar h_2)}(-1)^{h_1-\bar h_1}(2\pi)^2 \delta^{(2)}(z_1-z_2)\,,
\eadat\end{equation}
where $\sum_{i=1}^n h_i=\sum_{i=1}^n \bar h_i=2$, $h_i-\bar h_i \in \mathbb Z$.
Convergence of the integral requires $h_i +\bar h_i < 2$ for all $i$ although it may be extended by analytic continuation. 
}
\begin{equation}\label{ShTww}
\badat{2}
\tT_{\bw\bw}=12i\,Q^{0\,ren}_{2,\bw\bw}\,.
\eadat
\end{equation}
 From the point of view of a putative celestial CFT, the shadow transform thus puts superrotations and Diff($S^2$) shadow superrotations on equal footing as illustrated in Figure~\ref{fig:QshadowT}.

This relation elucidates the correspondence between the Ward identity for the Virasoro~\cite{Kapec:2014opa} and Diff($S^2$)~\cite{Campiglia:2014yka} symmetries and the subleading soft graviton theorem~\cite{Cachazo:2014fwa} as depicted in Figure~\ref{fig:ASG}. While the former corresponds to the insertion of the 2D stress tensor~\eqref{Tww} into the $\mathcal{S}$-matrix~\cite{Kapec:2016jld}, the latter corresponds to the insertion of the 2D shadow stress tensor~\eqref{ShTww}. Moreover, the relation between superrotations and the subleading soft graviton theorem becomes 1:1 when the asymptotic Virasoro symmetry group is enhanced to include shadow superrotations. 
We thus conclude that the asymptotic symmetry group of Einstein gravity at null infinity should include the closure of Virasoro under shadows within Diff$(S^2)$.  This extends to arbitrary diffeomorphisms $Y^z=Y^z(z,\bz)$ upon convolution of $\tT_{\bw\bw}$ with $D_w^3 Y^w(w,\bw)$, integrating by parts, and using $D_w^3 Y^z_{\bw\bw}=2\pi \delta^{(2)}(z-w)$.

\section*{Acknowledgements}
We would like to thank Adam Ball, Mina Himwich, Prahar Mitra, Romain Ruzziconi, Andy Strominger, Emilio Trevisani and Balt van Rees for valuable discussions. LD is supported by the European Union's Horizon 2020 research and innovation programme under the Marie Sk\l{}odowska-Curie grant agreement No.~746297. She also thanks the CPHT at Ecole Polytechnique for its hospitality. AP would like to thank the Centro de Ciencias de Benasque Pedro Pascual and the KITP Santa Barbara for its hospitality and acknowledges support from the National Science Foundation under Grant No. PHY-1748958. During the early stage of this project, SP was supported by the National Science Foundation through a Graduate Research Fellowship under grant DGE-1144152 and by the Hertz Foundation through a Harold and Ruth Newman Fellowship, and LD and AP were supported by the Black Hole Initiative at Harvard University, which is funded by a grant from the John Templeton Foundation.

\appendix

\section{Inner Product}\label{app:IP}
Here we evaluate the inner products (\ref{IPspin1Sigma}) and (\ref{IPspin2Sigma}) for two choices of Cauchy surfaces $\Sigma$, one at constant time $X^0=0$ and the other at null infinity.  For appropriate boundary conditions (at spatial infinity) one would expect the inner product to be the same for any Cauchy slice. The computations in this appendix show that this is indeed the case.

\subsection{Inner Product on $X^0=0$ Cauchy Slice}\label{app:IP1}
Let us start by considering the $X^0=0$ Cauchy slice. To effectively compute the inner products
 \begin{equation}\label{IPspin1app}
 (A, A')_{\Sigma_0} =- i \int d\Sigma^0 \,\left[ A^{\nu} (\nabla_0 {A'}_{\nu}^{*} -\nabla_\nu {A'}_{0}^{*} )
 -(A\leftrightarrow A'^*) \right] \,,
 \end{equation}
and
\begin{equation}\label{IPspin2app}
( h,h')_{\Sigma_0}=-i\int d\Sigma^0 \Big[ h^{\mu \nu} (\nabla_0 h'^{*}_{\,\,\mu \nu}-2 \nabla_\mu h'^{*}_{\,\,0\nu}) - (h \leftrightarrow h'^{*})\Big]\,,
\end{equation}
between two conformal primaries it is useful to reorganize the expressions~\eqref{ADelta} and~\eqref{hDelta}. Spin-one conformal primary can be written as
\begin{equation}\label{ADelta2}
A_{\mu;a}^{\Delta,\pm}(X^\mu;w,\bw)=\left[\p_a q_\mu+\frac{1}{\Delta} q_\mu \p_a\right]\frac{1}{(-q \cdot X_\pm)^{\Delta}}\,,
\end{equation}
and spin-two conformal primary as 
\begin{equation}\label{hDelta2}
h_{\mu\nu;a}^{\Delta,\pm}(X^\mu;w,\bw)
=\frac{1}{2}\left[\partial_{a} q_{\mu}\partial_{a} q_{\nu}+\frac{1}{\Delta}(q_\mu\p_a q_\nu+q_\nu\p_a q_\mu)\p_a +\frac{1}{\Delta(\Delta+1)} q_{\mu}q_\nu \p_a^2 \right] \frac{1}{(-q\cdot X_\pm )^{\Delta}}\,.
\end{equation}
where we have used $\p_a^2 q=0$. 
Using the Mellin representation for the denominator appearing in the conformal primaries~\eqref{fphi} in terms of plane waves
 \begin{equation}\label{eq:intone}
 \frac{(\pm i)^\Delta}{\Gamma(\Delta)} \int_0^\infty d\omega \omega^{\Delta-1}e^{\pm i\omega q\cdot X-\varepsilon\omega q^0}=\frac{1}{(-q\cdot X_\pm)^\Delta}\,,
 \end{equation}
one can see that the $d^3X$ integral of the product of two such terms -- parameterized by reference directions and weights $\{q,\Delta\}$ and $\{q',\Delta'\}$  -- should be proportional to a delta function on the celestial sphere $\delta^{(2)}(\hat{q}-\hat{q}')$,  where $\hat{q}$ refers to the unit vector pointing in the direction of the spatial part of $q$, forcing $\vec{q}$ and $\vec{q}'$ to be parallel. Indeed
 \begin{equation}\label{eq:scalardelta}
 \badat{2}
 I^\pm_1(\Delta,\Delta'^*+1)&\equiv\int d^3 X \frac{1}{(-q\cdot X_\pm )^{\Delta}(-q'\cdot X_\mp )^{\Delta'^*+1}}
 \\
  &= \frac{(2\pi)^3(\pm i)^{\Delta-\Delta'^*-1}}{2\Gamma(\Delta)\Gamma(\Delta'^*+1)}(q^0)^{-1}\delta^{(2)}(w-w')  \int_0^\infty d\omega \omega^{\Delta+\Delta'^*-3}e^{-2\varepsilon\omega q^0}\,.
\eadat
\end{equation}
Meanwhile, the $d^3 X$ integral gives 0 if both plane waves have the same $\pm$ sign.
Taking the limit $\varepsilon\rightarrow0$, one finds
\begin{equation}\label{eq:regdel}
\lim_{\varepsilon\rightarrow0}\int_0^\infty d\omega \omega^{\Delta+\Delta'^*-3}e^{-2\varepsilon\omega q^0}=2\pi\bdelta(i(\Delta+\Delta'^*-2)),
\end{equation}
which is the distribution we defined in section~\ref{subsec:gendelta}.  

To compute the inner products~\eqref{IPspin1app} and~\eqref{IPspin2app} for the primaries~\eqref{ADelta2} and~\eqref{hDelta2}, respectively, we use our above observation that we are restricted to the support of $\hat{q}=\hat{q}'$ and fact that the reference direction $q$ is a null vector satisfying
\begin{equation}
q\cdot q=q\cdot \p_a q=0\,,
\end{equation}
and similarly for $q'$. Furthermore, we have that $\p_a q \cdot \p_{a'^*}q =2\delta_{aa'}$.
For spin-one primaries we find
\begin{equation}\label{IPspin1Sigma20}
\badat{3}
(A_a^{\Delta,\pm}(w),A_{a'}^{\Delta',\pm}(w'))_{\Sigma_0} 
& = -i(\Delta'^*-1)q_0(\p_a q \cdot \p_{a'^*}q' )  I^\pm_1(\Delta,\Delta'^*+1) -(\Delta\leftrightarrow\Delta'^*)\\
 &=\pm 2(2\pi)^4e^{\pm i \pi \Delta}\sin(\Delta\pi) \frac{(\Delta-1)}{\pi\Delta(\Delta-2)}\delta_{aa'}\delta^{(2)}(w-w')\\
 &  \quad \times \bdelta(i(\Delta+\Delta'^*-2))\,,
 \eadat
\end{equation}
while for spin-two primaries
\begin{equation}\label{IPspin2Sigma20}
\badat{3}
(h_a^{\Delta,\pm}(w),h_{a'}^{\Delta',\pm}(w'))_{\Sigma_0} &=-\frac{i}{4}  (\Delta'^*-2)q_0 (\p_a q \cdot  \p_{a'^*} q')^2 I^\pm_1(\Delta,\Delta'^*+1)-(\Delta\leftrightarrow\Delta'^*)\\
 &= \pm (2\pi)^4 e^{\pm i \pi \Delta}\sin(\Delta\pi) \frac{(\Delta-1+i)(\Delta-1-i)}{\pi\Delta(\Delta-1)(\Delta-2)}\delta_{aa'}\delta^{(2)}(w-w')\\
 &  \quad \times \bdelta(i(\Delta+\Delta'^*-2))\,,
\eadat
\end{equation}
where we have taken the $\varepsilon\to0$ limit in the final expressions and made use of $\varphi(z)\bdelta(i(\Delta-z))\cong \varphi(\Delta)\bdelta(i(\Delta-z))$ which holds inside a contour integral of the kind~\eqref{eq:contour} since the conditions on $\varphi(z)$ of section~\ref{subsec:gendelta} are satisfied here. Also note that the exchanged terms involve $I_1^\pm(\Delta+1,\Delta'^*)$ and have the phase $(\pm i)^{\Delta+1-\Delta'^*}$ rather than $(\pm i)^{\Delta'^*-\Delta-1}$ due to the fact that the $i\varepsilon$ prescriptions are fixed for each mode.  

\subsection{Cauchy Slice Pushed to Null Infinity}\label{app:IP2}
To compute the inner product at null infinity, we can repeat the above steps once we evaluate the three-dimensional integral of two massless plane waves at future null infinity.
\begin{equation}
\badat{2}
 \lim_{r\rightarrow\infty}r^2\int d^2 z du  \sqrt{\gamma} e^{\mp i\omega q\cdot X}e^{\pm i\omega' q'\cdot X}&= \lim_{r\rightarrow\infty}r^2\int d^2\hat{x} du  e^{\pm i\omega q^0(u+r(1-\cos\theta))}e^{\mp i\omega'q'^0(u+r(1-\cos\theta'))}\\
 &=2\pi \delta(\omega q^0-\omega'q'^0) \lim_{r\rightarrow\infty}r^2\int d^2\hat{x}  e^{\pm i\omega q^0 r(1-\cos\theta)}e^{\mp i\omega' q'^0 r(1-\cos\theta')}\,,
\eadat
\end{equation}
where $\cos\theta=\hat{q}\cdot\hat{x}$ and $\cos\theta'=\hat{q}'\cdot\hat{x}$. Without loss of generality we can orient our axes so that $\hat{q}'$ points towards the north pole. Then using
\begin{equation}
\lim_{r\rightarrow\infty}\sin\theta' e^{i\omega' q'^0r(1-\cos\theta')}=\frac{i}{\omega' q'^0 r}\delta(\theta')+\mathcal{O}((\omega q^0 r)^{-2})\,,
\end{equation}
twice we have
\begin{equation}
\badat{2}
 \lim_{r\rightarrow\infty}r^2\int d^2 z du  \sqrt{\gamma}  e^{\mp i\omega q\cdot X}e^{\pm i\omega' q'\cdot X}&=
2\pi \delta(\omega q^0-\omega'q'^0) \lim_{r\rightarrow\infty}r^2[ \frac{\mp 2\pi i}{\omega q^0 r}  e^{\pm i\omega q^0 r(1-\hat{q}\cdot\hat{q}')}+\mathcal{O}((\omega q^0 r)^{-2})]\\
&=
(2\pi)^3 \delta(\omega q^0-\omega'q'^0)(\omega q^0)^{-2}\delta^{(2)}(\hat{q}-\hat{q}') \\
&=
(2\pi)^3 \delta^{(3)}(\omega \vec{q}-\omega' \vec{q}')\,.
\eadat
\end{equation}
Again we have used the notation $\hat{q}$ to refer to the unit vector pointing in the direction of the spatial part of $q$.  Since this agrees with the plane wave integral computed on the $X^0=0$ slice, we get the same result for  $I^\pm_1(\Delta,\Delta'^*+1)$ in~\eqref{eq:scalardelta}.  

The reasoning in Appendix~\ref{app:IP1} continues to hold except we must use $q\cdot n$ in place of $q_0$.  Now $n^\mu\p_\mu=\p_u-\frac{1}{2}\p_r$, so at the point near future null infinity parameterized by $\{u,r,z,\bz\}$ we have $q_u=q_0=-(1+w\bw)$ and $q_r=-\frac{2|z-w|^2}{1+z\bz}$.  Because of the $z$ dependence of $q_r$, unlike in the spatial case, any $q_r$ term modifies the Cauchy slice integrand. However, since we saw above that the Cauchy slice integral localizes to $z=w$ any term with $q_r$ or $\p_a q_r$ will vanish.  We thus find that our answer at future null infinity is the same as our computation at the $X^0=0$ slice.

\section{Harmonic Diff($S^2$) Vector Field}\label{app:DiffS2}
In this appendix we derive the vector field $\xi$ that generates arbitrary diffeomorphisms of the celestial sphere in harmonic gauge where
\begin{equation}\label{eq:harmonic}
\Box \xi_\mu(u,r,z,\bz)=0\,.
\end{equation}
Our ansatz for the radial expansion of the vector field is\footnote{Here $\sim$ denotes logarithmic terms in the expansion and is unrelated to the $\sim$ we use for the shadow transform. For now, we will let $n$ range over all integers, but boundary conditions will demand that the coefficients $\{\xi_\mu^{(n)},\tilde{\xi}_\mu^{(n)}\}$ are non-zero only for a semi-infinite range extending to $n\rightarrow+\infty$.  The harmonic and residual gauge fixing conditions provide restrictions on the leading data.} 
\begin{equation}
\xi_\mu(u,r,z,\bz)=\sum_n r^{-n}(\xi_\mu^{(n)}(u,z,\bz)+\log(r) \,\tilde{\xi}_\mu^{(n)}(u,z,\bz))\,,
\end{equation}
which, after a similar decomposition of~\eqref{eq:harmonic} into powers of $r^{-n}$ yields~\cite{Himwich:2019qmj}
\begin{equation}
\badat{6}\label{eq:boxxi}
[\square \xi_u]^{(n)} &= 2(n-2)\partial_u \xi_u^{(n-1)} + \left[D^2 +(n-2)(n-3) \right]\xi_u^{(n-2)}\\
&~~~+ (5-2n)\tilde{\xi}_u^{(n-2)} - 2\partial_u\tilde{\xi}_u^{(n-1)}\,, \\
[\square \xi_r]^{(n)} &= 2(n-2)\partial_u \xi_r^{(n-1)} + \left[D^2 +(n-2)(n-3) - 2\right]\xi_r^{(n-2)}+ 2 \xi_u^{(n-2)}- 2 D^A\xi_A^{(n-3)} \\
&~~~+ (5-2n)\tilde{\xi}_r^{(n-2)} - 2\partial_u\tilde{\xi}_r^{(n-1)}\,, \\
[\square \xi_A]^{(n-1)} &= 2(n-2)\partial_u \xi_A^{(n-2)} + \left[D^2 +(n-2)(n-3) -1 \right]\xi_A^{(n-3)}- 2 \partial_A\left(\xi_u^{(n-2)} - \xi_r^{(n-2)}\right) \\
&~~~+ (5-2n)\tilde{\xi}_A^{(n-3)} - 2\partial_u\tilde{\xi}_A^{(n-2)} \,,
\eadat
\end{equation}
and at logarithmic order 
\begin{equation}
\badat{3}\label{eq:xilog}
&[\square \tilde{\xi}_u]^{(n)} = 2(n-2)\partial_u \tilde{\xi}_u^{(n-1)} + \left[D^2 +(n-2)(n-3) \right]\tilde{\xi}_u^{(n-2)}\,, \\
&[\square \tilde{\xi}_r]^{(n)} = 2(n-2)\partial_u \tilde{\xi}_r^{(n-1)} + \left[D^2 +(n-2)(n-3) - 2\right]\tilde{\xi}_r^{(n-2)} + 2 \tilde{\xi}_u^{(n-2)} - 2 D^A\tilde{\xi}_A^{(n-3)}\,, \\
&[\square \tilde{\xi}_A]^{(n-1)} = 2(n-2)\partial_u \tilde{\xi}_A^{(n-2)} + \left[D^2 +(n-2)(n-3) -1 \right]\tilde{\xi}_A^{(n-3)} - 2 \partial_A\left(\tilde{\xi}_u^{(n-2)} - \tilde{\xi}_r^{(n-2)}\right) \,.
\eadat
\end{equation}
Unlike in~\cite{Himwich:2019qmj}, we use a radial gauge condition to perform our residual gauge fixing.  This takes the form
\begin{equation}
V_\nu\equiv X^\mu(\nabla_\mu \xi_\nu+\nabla_\nu \xi_\mu)=0\,,
\end{equation}
which decomposes into 
\begin{equation}\label{eq:resfix1}
 \badat{3}
&V_u^{(n)}
=-2u\p_u\xi^{u(n)}-2u\p_u\xi^{r(n)}-\p_u\xi^{u(n+1)}+n\xi^{u(n)}+n\xi^{r(n)}-\tilde{\xi}^{u(n)}-\tilde{\xi}^{r(n)}\,,\\
&V_r^{(n)}
=-u\p_u\xi^{u(n)}+(n-1)u\xi^{u(n-1)}+(n-1)u\xi^{r(n-1)}-u\tilde{\xi}^{u(n-1)}-u\tilde{\xi}^{r(n-1)}+2n\xi^{u(n)}-2\tilde{\xi}^{u(n)}\,,\\
&V_A^{(n-1)}
=\gamma_{AB}u\p_u\xi^{B(n+1)}-u\p_A\xi^{u(n-1)}-u\p_A\xi^{r(n-1)}-\p_A\xi^{u(n)}-(n+1)\gamma_{AB}\xi^{B(n+1)}+\gamma_{AB}\tilde{\xi}^{B(n+1)}\,,
\eadat
\end{equation}
and for the logarithmic terms into
\begin{equation}\label{eq:resfix2}
 \badat{3}
&\tilde{V}_u^{(n)}
=-2u\p_u\tilde{\xi}^{u(n)}-2u\p_u\tilde{\xi}^{r(n)}-\p_u\tilde{\xi}^{u(n+1)}+n\tilde{\xi}^{u(n)}+n\tilde{\xi}^{r(n)}\,,\\
&\tilde{V}_r^{(n)}
=-u\p_u\tilde{\xi}^{u(-1)}-2u\tilde{\xi}^{u(-2)}-2u\tilde{\xi}^{r(-2)}+2n\tilde{\xi}^{u(-1)}\,,\\
&\tilde{V}_A^{(n-1)}
=\gamma_{AB}u\p_u\tilde{\xi}^{B(n+1)}-u\p_A\tilde{\xi}^{u(n-1)}-u\p_A\tilde{\xi}^{r(n-1)}-\p_A\tilde{\xi}^{u(n)}-(n+1)\gamma_{AB}\tilde{\xi}^{B(n+1)}\,.
\eadat
\end{equation}
We use $\xi^u=-\xi_r$, $\xi^r=\xi_r-\xi_u$ to combine the above expressions involving raised vs lowered indexed components.

In this appendix we are interested in the specific example of a  Diff$(S^2)$ transformation of the celestial sphere, i.e.
\begin{equation}
\xi^{A(0)}=Y^A(z,\bz)\,,
\end{equation}
which corresponds in our notation to 
\begin{equation}
\xi_A^{(-2)}=Y_A(z,\bz)\,.
\end{equation}
Even if we only cared about pure superrotations, for which $Y^z(z,\bz)=Y^z(z)$, we would need the full Diff$(S^2)$ vector field in order to keep track of contact terms.\\
We start by imposing the conditions
\begin{equation}
\tilde{\xi}_u^{(n\le 0)}=\tilde{\xi}_r^{(n\le 0)}=\tilde{\xi}_A^{(n\le -1)}=0\,, \quad {\xi}_A^{(n\le -3)}=0\, ,\quad{\xi}_A^{(-2)}=Y_A(z,\bz)\,,
\end{equation}
which is reasoned as follows.  The last two conditions imply a Diff$(S^2)$ on the celestial sphere, while the first three conditions follow from consistency of the log-order equations.  The first non-zero orders allowed would be the free data of the log-equations, but are fixed by the non-log equations.

The $u$-independence of $Y^A(z,\bz)$ gives
\begin{equation}\label{eq:radialgaugecon}
 \badat{3}
&V_u^{(n\le-1)}
=(-2u\p_u+n)(\xi^{u(n)}+\xi^{r(n)})-\p_u\xi^{u(n+1)}\,,\\
&V_r^{(n+1\le0)}
=(-u\p_u+2n+2)\xi^{u(n+1)}+nu(\xi^{u(n)}+\xi^{r(n)})\,,\\
&V_A^{(n-1\le-2)}
=-\p_A(u(\xi^{u(n)}+\xi^{r(n)})+\xi^{u(n+1)})\,.
\eadat
\end{equation}
Noting that $\xi^u+\xi^r=-\xi_u$, the radial gauge condition~\eqref{eq:radialgaugecon} tells us 
\begin{equation}
\xi_r^{(n+1)}=-u \xi_u^{(n)}\,,
\end{equation}
for $n\le -1$.
We can then use this to replace every $\xi_r$ in~\eqref{eq:boxxi}, in particular 
\begin{equation}
[\square \xi_A]^{(n-1\le-1)} =  2 \partial_A\left(\xi_u^{(n-2)} - \xi_r^{(n-2)}\right)\,, 
\ee
then implies \be
\xi_u^{(n-2)} =-u \xi_u^{(n-3)}\,,
\end{equation}
for $n\le0$.  If we want the tower to truncate we should demand
\begin{equation}
\xi_u^{(n\le-2)}=0\,.
\end{equation}
We then have the leading equations
\begin{equation}\label{eq:mainsolve}
\badat{7}
[\square \xi_A]^{(0)} &= -2\partial_u \xi_A^{(-1)} + \left[D^2 +1 \right]\xi_A^{(-2)}- 2 \partial_A \xi_u^{(-1)}\,, \\
[\square \xi_r]^{(1)} &= 2u\partial_u \xi_u^{(-1)} + 4 \xi_u^{(-1)}- 2 D^A\xi_A^{(-2)}\,, \\
[\square \xi_u]^{(0)} &= -4\partial_u \xi_u^{(-1)}\,, \\
[\square \xi_u]^{(1)} &= -2\partial_u \xi_u^{(0)} + \left[D^2 +2 \right]\xi_u^{(-1)}\,,\\
[\square \xi_A]^{(1)} &=  \left[D^2 -1 \right]\xi_A^{(-1)}- 2 \partial_A\left(\xi_u^{(0)} +u\xi_u^{(-1)}\right)- 2\partial_u\tilde{\xi}_A^{(0)}\,, \\
[\square \xi_r]^{(2)} &=  -\left[D^2  - 2\right]u\xi_u^{(-1)}+ 2 \xi_u^{(0)}- 2 D^A\xi_A^{(-1)} - 2\partial_u\tilde{\xi}_r^{(1)}\,, \\
[\square \xi_u]^{(2)} &=  D^2 \xi_u^{(0)}- 2\partial_u\tilde{\xi}_u^{(1)}\,.
\eadat
\end{equation}
The first three tell us
\begin{equation}
\badat{2}
& \xi_u^{(-1)}=\frac{1}{2}D^AY_A\,, \quad \xi_u^{(0)}=\frac{u}{4}\left[D^2 +2 \right] D_AY^A\,, \quad\xi_r^{(0)}=-\frac{u}{2}D^A\xi_A^{(-2)}\,,\\
&\xi_A^{(-2)}=Y_A(z,\bz)\,, \quad \xi_A^{(-1)} =\frac{u}{2} (\left[D^2 +1 \right]Y_A-  D_A D_B Y^B )\,,
\eadat
\end{equation}
so that
\begin{equation}\label{eq:diffs2form}
\badat{3}
\xi^A&=Y^A+\frac{u}{2r}(\left[D^2 +1 \right]Y^A-  D^A D_B Y^B)+\dots\,,\\
\xi^u&=\frac{u}{2} D^A Y_A+\dots\,,\\
\xi^r&=-\frac{r}{2} D^A Y_A-\frac{u}{4}\left[D^2 +4 \right]D^A Y_A+\dots\,,
\eadat
\end{equation}
where the omitted terms should be solved for using the radial gauge fixing recursions~\eqref{eq:resfix1}-\eqref{eq:resfix2}, consistent with the harmonic condition~\eqref{eq:harmonic}. 

Meanwhile the last three equations in~\eqref{eq:mainsolve} allow us solve for the leading log terms, which are in turn free data for the log tower~\eqref{eq:xilog}.  Explicitly, these are
\begin{equation}
\badat{3}
 2\partial_u\tilde{\xi}_A^{(0)} & =  \frac{u}{2}(\left[D^2 -1 \right] \left[D^2 +1 \right]Y_A - 2 D_A \left[D^2 +2 \right] D_BY^B  )\,,  \\
2\partial_u\tilde{\xi}_r^{(1)}&=  \frac{u}{2}(2\left[D^2  + 2\right]D^AY_A- 2 D^A \left[D^2 +1 \right]Y_A )\,, \\ 
2\partial_u\tilde{\xi}_u^{(1)} &=  \frac{u}{4}D^2\left[D^2 +2 \right] D_AY^A\,. \\
\eadat
\end{equation}
Finally, we note that away from poles of meromorphic $Y^z(z)$, we have
\begin{equation}\label{eq:contactreally}
[D^2+1]Y^A=0\,, \quad [D^2+2]D_AY^A=0\,.
\end{equation}
We thus see that all of the log terms vanish, while the vector field reduces to
\begin{equation}\label{finalxiup}
\badat{3}
\xi^A&=Y^A-\frac{u}{2r} D^A D_B Y^B+\dots\,,\\
\xi^u&=\frac{u}{2} D^A Y_A+\dots\,,\\
\xi^r&=-\frac{r}{2} D^A Y_A-\frac{u}{2}D^A Y_A+\dots\,.
\eadat
\end{equation}
It should be emphasized, that in order to keep track of poles in $Y^z$, as needed in section~\ref{sec:CSMem}, we should use the generic Diff$(S^2)$ form~\eqref{eq:diffs2form} which adds contact term corrections to~\eqref{finalxiup}.

\section{Renormalized Symplectic Structure}\label{app:Qspin2}
In this appendix, we compute the inner product at future null infinity for the conformal primary Goldstone modes with conformal dimension $\Delta=0$ and $\tilde \Delta=2$. This amounts to computing the Iyer-Wald symplectic structure~\cite{Iyer:1994ys,Lee:1990nz} 
\begin{equation}
\Omega[\delta g, \delta'g;g]=\int_{\scri^+} \omega[\delta g, \delta' g;g]\,,
\end{equation}
for the variations $h=\delta g$, $h'=\delta' g$ around a fixed background $g$, where the presymplectic form $\omega=\omega^\rho (d^3x)_\rho$ is given by
\begin{equation}\label{omrho2}
\omega^\rho[\delta g, \delta' g;g]= r^2\sqrt{\gamma}(h^{\mu \nu} \nabla^\rho h'_{\,\,\mu \nu}-2h^{\mu \nu} \nabla_\mu h'^\rho_{\,\,\,\nu}- (h \leftrightarrow h'))\,.
\end{equation}
We want to compute~\eqref{omrho2} for $h'_{\mu\nu}=h^0_{\mu\nu;\bw\bw}$ and $h'_{\mu\nu}=\th^2_{\mu\nu;ww}$, both of which behave near future null infinity as\footnote{The superscript $(n)$ denotes the coefficient of the $1/r^n$ term in the expansion near null infinity.}
\begin{equation}\label{h'}
\badat{2}
& h'_{uu}=\mathcal O(r^{0}) \virg h'_{uz}=\mathcal O(r^0)\, \virg h'_{u\bz}=r \delta'{g}^{(-1)}_{u\bz}+\mathcal O(r^0)\,, \\
&h'_{zz}=r \delta' C_{zz}+\mathcal O(r^{0}) \virg h'_{z\bz}=r \delta' C_{z\bz}+\mathcal O(r^{0})\virg h'_{\bz\bz}=r^2\delta' \gamma_{\bz \bz}+r \delta' C_{\bz\bz}+\mathcal O(r^{0})\,,\\
\eadat
\end{equation}
where the variations $\delta' C_{AB}$ depend on $(u,z,\bz)$, while $\delta'{g}^{(-1)}_{u\bz}$ and $\delta' \gamma_{\bz \bz}$ only depend on the angles.
We assume the following fall-offs for the perturbation $h_{\mu\nu}=\delta g_{\mu \nu}$:
\begin{equation}\label{h}
h_{uu}=\delta g_{uu}^{(0)}+\mathcal O(r^{-1})\,, \quad h_{uA}=\delta g_{uA}^{(0)}+\mathcal O(r^{-1})\,, \quad h_{AB}=r \delta C_{AB}+\mathcal O(r^{0}),
\end{equation}
with $\delta C_{z\bz}=0$ and all variations are functions of $(u,z,\bz)$. The leading behavior of the radial metric components in both~\eqref{h'} and~\eqref{h} is determined by the radial gauge condition 
\begin{equation}
h_{r\mu}=-\frac{u}{r}h_{u\mu}\,.
\end{equation}
Evaluating~\eqref{omrho2} for these fall-offs, we find:
\begin{equation}
\badat{2}
&\omega^{u}=- \sqrt \gamma \delta'\gamma_{\bz \bz}\delta C^{\bz \bz}+\mathcal O(r^{-1}),\\
&\omega^{r}=r  \sqrt \gamma \delta'\gamma_{\bz \bz}\delta N^{\bz \bz}+\omega^{r(0)}+\mathcal O(r^{-1}),
\eadat
\end{equation} 
with $N_{AB}=\p_u C_{AB}$ and
\begin{equation}\label{IPspin2BondiData}
\badat{2}
\omega^{r(0)}=&
 \sqrt \gamma \,\Big( \delta' C_{AB}\delta N^{AB}-\delta C_{AB}\delta' N^{AB}+\delta'\gamma^{zz}(2D_z \delta g_{uz}^{(0)}-\partial_u \delta g_{zz}^{(0)})\\
 &\quad \quad+2D_\bz \delta' g_{u\bz}^{(-1)}\delta C^{\bz \bz}
  - 2\gamma^{z\bz} \delta'{g}^{(-1)}_{u\bz}(1-u\partial_u) \delta g_{uz}^{(0)}\Big).
\eadat
\end{equation} 
Similar expressions can be obtained for the opposite helicity Goldstone modes $h'_{\mu\nu}=h^0_{\mu\nu;ww}$ and $h'_{\mu\nu}=\th^2_{\mu\nu;\bw\bw}$ whose asymptotic behavior is given by~\eqref{h'} with $z$ and $\bz$ exchanged.

We thus see that the presymplectic form diverges in $r$, a feature which was already observed in~\cite{Compere:2018ylh}. This divergence can be removed via an appropriate renormalization of $\omega$. Writing the presymplectic form in terms of the presymplectic potential $\Theta$,
\begin{equation}
\omega[\delta g, \delta' g;g]=\delta \Theta[\delta' g; g]-\delta' \Theta[\delta g; g]\,,
\end{equation}
exposes the freedom to perform a shift\footnote{This renormalization procedure was recently shown to be covariant in terms of the boundary structure, justifying a posteriori the counterterm prescription to remove the radial divergence \cite{Compere:2020lrt}.}
\begin{equation}\label{amb}
\Theta \to \Theta +dY\,,
\end{equation}
where $Y$ is a spacetime co-dimension 2 form. 
The freedom~\eqref{amb} expresses the residual ambiguity in the definition of the presymplectic potential in the covariant phase space formalism~\cite{Iyer:1994ys}. We will adopt the same choice of $Y$ as given in~\cite{Compere:2018ylh} and will not repeat the details here. As shown there, the effect of this renormalization is to remove the radially divergent piece in $\omega^r$, as well as remove the finite piece in $\omega^u$.
The renormalized inner product therefore only receives contributions from the finite piece in $\omega^r$:
\begin{equation}
\Omega^{ren}[\delta g, \delta' g;g]=\int du d^2 z \, \omega^{r(0)}[\delta g, \delta' g;g]\,.
\end{equation}
To simplify~\eqref{IPspin2BondiData} we make use of the equations of motion (see e.g. Appendix~A of~\cite{Himwich:2019qmj})
\begin{equation}
\badat{3}
0=\Box h_{rz} \quad &\to \quad (1-u\partial_u) \delta g^{(0)}_{uz}=D^z \delta C_{zz}\,,\\
0=\Box h_{zz} \quad &\to \quad 4D_z \delta g^{(0)}_{uz}=(D^2-2)\delta C_{zz}+2\partial_u \delta g^{(0)}_{zz}\,,\\
0=\Box h'_{\bz\bz} \quad &\to \quad 4D_\bz \delta' g^{(-1)}_{u\bz}=(D^2-2)\delta' \gamma_{\bz\bz}\,,
\eadat
\end{equation}
and integrate by parts on the celestial sphere. This yields
\begin{equation}\label{wr}
\badat{2}
\omega^{r(0)}&=
 \sqrt \gamma ( \delta' C_{AB}\delta N^{AB}-\delta C_{AB}\delta' N^{AB}+\frac{1}{2}D^2\delta'\gamma_{\bz\bz}\delta C^{\bz \bz}).
\eadat
\end{equation}
Finally, we can express the inner product in terms of the vector field $Y^A=(Y^z,Y^\bz=0)$ using that the Goldstone modes are pure shift transformations $\delta'=\delta_Y^{shift}$ (see \eqref{tC} and \eqref{N0}), namely
\begin{equation}
\badat{2}
&\delta' C_{zz}=-uD_z^3 Y^z\,, \quad \delta' C_{\bz\bz}=\frac{u}{2}D^2 D_\bz Y_\bz\,, \quad \delta' \gamma_{\bz\bz} = 2 D_\bz Y_\bz\,,\\
&\delta' N_{zz}=-D_z^3 Y^z\,,\quad \delta' N_{\bz\bz}=\frac{1}{2}D^2 D_\bz Y_\bz\,,
\eadat
\end{equation}
where we used the identities
\begin{equation}
D_\bz D^\bz Y_\bz=\frac{1}{2}(D^2 -1)Y_\bz\,, \quad [D_\bz, D^2]Y_\bz=-3D_\bz Y_\bz\,.
\end{equation}
We thus find
\begin{equation}\label{omDsoft}
\badat{2}
\Omega^{ren}[\delta g,\delta_{Y}^{\mathrm {shift}} g;g]=\int du d^2 z  \sqrt \gamma \, (D_z^3 Y^z\delta C^{zz}+\frac{1}{2}D^2 D_\bz Y_\bz \delta C^{\bz \bz}-u D_z^3 Y^z \delta N^{zz}+\frac{u}{2}D^2 D_\bz Y_\bz \delta N^{\bz \bz}).
\eadat
\end{equation}
Everything being linear in the metric fields, one can integrate out the variation as $\Omega^{ren}[\delta g,\delta_{Y}^{\mathrm {shift}} g;g]\equiv-\delta \mathcal Q_Y $, with
\begin{equation}
\badat{2}
\mathcal Q_Y=\int du d^2 z  \sqrt \gamma \, (-D_z^3 Y^z  C^{zz}-\frac{1}{2}D^2 D_\bz Y_\bz  C^{\bz \bz}+u D_z^3 Y^z  N^{zz}-\frac{u}{2}D^2 D_\bz Y_\bz  N^{\bz \bz}).
\eadat
\end{equation}

\bibliographystyle{utphys}
\bibliography{references}

\providecommand{\href}[2]{#2}\begingroup\raggedright\begin{thebibliography}{10}

\bibitem{Strominger:2013lka}
A.~Strominger, ``{Asymptotic Symmetries of Yang-Mills Theory},''
  \href{http://dx.doi.org/10.1007/JHEP07(2014)151}{{\em JHEP} {\bfseries 07}
  (2014) 151},
\href{http://arxiv.org/abs/1308.0589}{{\ttfamily arXiv:1308.0589 [hep-th]}}.

\bibitem{Strominger:2013jfa}
A.~Strominger, ``{On BMS Invariance of Gravitational Scattering},''
  \href{http://dx.doi.org/10.1007/JHEP07(2014)152}{{\em JHEP} {\bfseries 07}
  (2014) 152},
\href{http://arxiv.org/abs/1312.2229}{{\ttfamily arXiv:1312.2229 [hep-th]}}.

\bibitem{Strominger:2017zoo}
A.~Strominger, {\em {Lectures on the Infrared Structure of Gravity and Gauge
  Theory}}.
\newblock {Princeton University Press}, 2018.
\newblock
\href{http://arxiv.org/abs/1703.05448}{{\ttfamily arXiv:1703.05448 [hep-th]}}.
\newblock

\bibitem{deBoer:2003vf}
J.~de~Boer and S.~N. Solodukhin, ``{A Holographic reduction of Minkowski
  space-time},'' \href{http://dx.doi.org/10.1016/S0550-3213(03)00494-2}{{\em
  Nucl. Phys.} {\bfseries B665} (2003) 545--593},
\href{http://arxiv.org/abs/hep-th/0303006}{{\ttfamily arXiv:hep-th/0303006
  [hep-th]}}.

\bibitem{Bondi:1962px}
H.~Bondi, M.~G.~J. van~der Burg, and A.~W.~K. Metzner, ``{Gravitational waves
  in general relativity. 7. Waves from axisymmetric isolated systems},''
\href{http://dx.doi.org/10.1098/rspa.1962.0161}{{\em Proc. Roy. Soc. Lond.}
  {\bfseries A269} (1962) 21--52}.

\bibitem{Sachs:1962wk}
R.~K. Sachs, ``{Gravitational waves in general relativity. 8. Waves in
  asymptotically flat space-times},''
\href{http://dx.doi.org/10.1098/rspa.1962.0206}{{\em Proc. Roy. Soc. Lond.}
  {\bfseries A270} (1962) 103--126}.

\bibitem{Sachs:1962zza}
R.~Sachs, ``{Asymptotic symmetries in gravitational theory},''
\href{http://dx.doi.org/10.1103/PhysRev.128.2851}{{\em Phys. Rev.} {\bfseries
  128} (1962) 2851--2864}.

\bibitem{Kapec:2014opa}
D.~Kapec, V.~Lysov, S.~Pasterski, and A.~Strominger, ``{Semiclassical Virasoro
  symmetry of the quantum gravity $ \mathcal{S}$-matrix},''
  \href{http://dx.doi.org/10.1007/JHEP08(2014)058}{{\em JHEP} {\bfseries 08}
  (2014) 058},
\href{http://arxiv.org/abs/1406.3312}{{\ttfamily arXiv:1406.3312 [hep-th]}}.

\bibitem{Banks:2003vp}
T.~Banks, ``{A Critique of pure string theory: Heterodox opinions of diverse
  dimensions},'' \href{http://arxiv.org/abs/hep-th/0306074}{{\ttfamily
  arXiv:hep-th/0306074}}.

\bibitem{Barnich:2009se}
G.~Barnich and C.~Troessaert, ``{Symmetries of asymptotically flat 4
  dimensional spacetimes at null infinity revisited},''
  \href{http://dx.doi.org/10.1103/PhysRevLett.105.111103}{{\em Phys. Rev.
  Lett.} {\bfseries 105} (2010) 111103},
\href{http://arxiv.org/abs/0909.2617}{{\ttfamily arXiv:0909.2617 [gr-qc]}}.

\bibitem{Barnich:2011ct}
G.~Barnich and C.~Troessaert, ``{Supertranslations call for superrotations},''
  {\em PoS} {\bfseries CNCFG} (2010) 010,
  \href{http://arxiv.org/abs/1102.4632}{{\ttfamily arXiv:1102.4632 [gr-qc]}}.
[Ann. U. Craiova Phys.21,S11(2011)].

\bibitem{Barnich:2010eb}
G.~Barnich and C.~Troessaert, ``{Aspects of the BMS/CFT correspondence},''
  \href{http://dx.doi.org/10.1007/JHEP05(2010)062}{{\em JHEP} {\bfseries 05}
  (2010) 062},
\href{http://arxiv.org/abs/1001.1541}{{\ttfamily arXiv:1001.1541 [hep-th]}}.

\bibitem{Cachazo:2014fwa}
F.~Cachazo and A.~Strominger, ``{Evidence for a New Soft Graviton Theorem},''
\href{http://arxiv.org/abs/1404.4091}{{\ttfamily arXiv:1404.4091 [hep-th]}}.

\bibitem{Kapec:2016jld}
D.~Kapec, P.~Mitra, A.-M. Raclariu, and A.~Strominger, ``{2D Stress Tensor for
  4D Gravity},'' \href{http://dx.doi.org/10.1103/PhysRevLett.119.121601}{{\em
  Phys. Rev. Lett.} {\bfseries 119} no.~12, (2017) 121601},
\href{http://arxiv.org/abs/1609.00282}{{\ttfamily arXiv:1609.00282 [hep-th]}}.

\bibitem{Pasterski:2016qvg}
S.~Pasterski, S.-H. Shao, and A.~Strominger, ``{Flat Space Amplitudes and
  Conformal Symmetry of the Celestial Sphere},''
  \href{http://dx.doi.org/10.1103/PhysRevD.96.065026}{{\em Phys. Rev.}
  {\bfseries D96} no.~6, (2017) 065026},
\href{http://arxiv.org/abs/1701.00049}{{\ttfamily arXiv:1701.00049 [hep-th]}}.

\bibitem{Pasterski:2017kqt}
S.~Pasterski and S.-H. Shao, ``{Conformal basis for flat space amplitudes},''
  \href{http://dx.doi.org/10.1103/PhysRevD.96.065022}{{\em Phys. Rev.}
  {\bfseries D96} no.~6, (2017) 065022},
\href{http://arxiv.org/abs/1705.01027}{{\ttfamily arXiv:1705.01027 [hep-th]}}.

\bibitem{Pasterski:2017ylz}
S.~Pasterski, S.-H. Shao, and A.~Strominger, ``{Gluon Amplitudes as 2d
  Conformal Correlators},''
  \href{http://dx.doi.org/10.1103/PhysRevD.96.085006}{{\em Phys. Rev.}
  {\bfseries D96} no.~8, (2017) 085006},
\href{http://arxiv.org/abs/1706.03917}{{\ttfamily arXiv:1706.03917 [hep-th]}}.

\bibitem{Cheung:2016iub}
C.~Cheung, A.~de~la Fuente, and R.~Sundrum, ``{4D scattering amplitudes and
  asymptotic symmetries from 2D CFT},''
  \href{http://dx.doi.org/10.1007/JHEP01(2017)112}{{\em JHEP} {\bfseries 01}
  (2017) 112},
\href{http://arxiv.org/abs/1609.00732}{{\ttfamily arXiv:1609.00732 [hep-th]}}.

\bibitem{Donnay:2018neh}
L.~Donnay, A.~Puhm, and A.~Strominger, ``{Conformally Soft Photons and
  Gravitons},'' \href{http://dx.doi.org/10.1007/JHEP01(2019)184}{{\em JHEP}
  {\bfseries 01} (2019) 184},
\href{http://arxiv.org/abs/1810.05219}{{\ttfamily arXiv:1810.05219 [hep-th]}}.

\bibitem{Fan:2019emx}
W.~Fan, A.~Fotopoulos, and T.~R. Taylor, ``{Soft Limits of Yang-Mills
  Amplitudes and Conformal Correlators},''
  \href{http://dx.doi.org/10.1007/JHEP05(2019)121}{{\em JHEP} {\bfseries 05}
  (2019) 121},
\href{http://arxiv.org/abs/1903.01676}{{\ttfamily arXiv:1903.01676 [hep-th]}}.

\bibitem{Pate:2019mfs}
M.~Pate, A.-M. Raclariu, and A.~Strominger, ``{Conformally Soft Theorem in
  Gauge Theory},'' \href{http://dx.doi.org/10.1103/PhysRevD.100.085017}{{\em
  Phys. Rev.} {\bfseries D100} no.~8, (2019) 085017},
\href{http://arxiv.org/abs/1904.10831}{{\ttfamily arXiv:1904.10831 [hep-th]}}.

\bibitem{Adamo:2019ipt}
T.~Adamo, L.~Mason, and A.~Sharma, ``{Celestial amplitudes and conformal soft
  theorems},'' \href{http://dx.doi.org/10.1088/1361-6382/ab42ce}{{\em Class.
  Quant. Grav.} {\bfseries 36} no.~20, (2019) 205018},
  \href{http://arxiv.org/abs/1905.09224}{{\ttfamily arXiv:1905.09224
  [hep-th]}}.

\bibitem{Puhm:2019zbl}
A.~Puhm, ``{Conformally Soft Theorem in Gravity},''
  \href{http://arxiv.org/abs/1905.09799}{{\ttfamily arXiv:1905.09799
  [hep-th]}}.

\bibitem{Guevara:2019ypd}
A.~Guevara, ``{Notes on Conformal Soft Theorems and Recursion Relations in
  Gravity},''
\href{http://arxiv.org/abs/1906.07810}{{\ttfamily arXiv:1906.07810 [hep-th]}}.

\bibitem{Law:2019glh}
Y.~A. Law and M.~Zlotnikov, ``{Poincar\'e constraints on celestial
  amplitudes},'' \href{http://dx.doi.org/10.1007/JHEP03(2020)085}{{\em JHEP}
  {\bfseries 20} (2020) 085}, \href{http://arxiv.org/abs/1910.04356}{{\ttfamily
  arXiv:1910.04356 [hep-th]}}.

\bibitem{Fotopoulos:2019vac}
A.~Fotopoulos, S.~Stieberger, T.~R. Taylor, and B.~Zhu, ``{Extended BMS Algebra
  of Celestial CFT},'' \href{http://dx.doi.org/10.1007/JHEP03(2020)130}{{\em
  JHEP} {\bfseries 03} (2020) 130},
  \href{http://arxiv.org/abs/1912.10973}{{\ttfamily arXiv:1912.10973
  [hep-th]}}.

\bibitem{Banerjee:2020kaa}
S.~Banerjee, S.~Ghosh, and R.~Gonzo, ``{BMS symmetry of celestial OPE},''
  \href{http://dx.doi.org/10.1007/JHEP04(2020)130}{{\em JHEP} {\bfseries 04}
  (2020) 130}, \href{http://arxiv.org/abs/2002.00975}{{\ttfamily
  arXiv:2002.00975 [hep-th]}}.

\bibitem{Campiglia:2015yka}
M.~Campiglia and A.~Laddha, ``{New symmetries for the Gravitational
  S-matrix},'' \href{http://dx.doi.org/10.1007/JHEP04(2015)076}{{\em JHEP}
  {\bfseries 04} (2015) 076},
\href{http://arxiv.org/abs/1502.02318}{{\ttfamily arXiv:1502.02318 [hep-th]}}.

\bibitem{Compere:2018ylh}
G.~Comp\`ere, A.~Fiorucci, and R.~Ruzziconi, ``{Superboost transitions,
  refraction memory and super-Lorentz charge algebra},''
  \href{http://dx.doi.org/10.1007/JHEP11(2018)200}{{\em JHEP} {\bfseries 11}
  (2018) 200}, \href{http://arxiv.org/abs/1810.00377}{{\ttfamily
  arXiv:1810.00377 [hep-th]}}.

\bibitem{Campiglia:2020qvc}
M.~Campiglia and J.~Peraza, ``{Generalized BMS charge algebra},''
  \href{http://arxiv.org/abs/2002.06691}{{\ttfamily arXiv:2002.06691 [gr-qc]}}.

\bibitem{Campiglia:2014yka}
M.~Campiglia and A.~Laddha, ``{Asymptotic symmetries and subleading soft
  graviton theorem},'' \href{http://dx.doi.org/10.1103/PhysRevD.90.124028}{{\em
  Phys. Rev.} {\bfseries D90} no.~12, (2014) 124028},
\href{http://arxiv.org/abs/1408.2228}{{\ttfamily arXiv:1408.2228 [hep-th]}}.

\bibitem{Nandan:2019jas}
D.~Nandan, A.~Schreiber, A.~Volovich, and M.~Zlotnikov, ``{Celestial
  Amplitudes: Conformal Partial Waves and Soft Limits},''
  \href{http://dx.doi.org/10.1007/JHEP10(2019)018}{{\em JHEP} {\bfseries 10}
  (2019) 018},
\href{http://arxiv.org/abs/1904.10940}{{\ttfamily arXiv:1904.10940 [hep-th]}}.

\bibitem{Pate:2019lpp}
M.~Pate, A.-M. Raclariu, A.~Strominger, and E.~Y. Yuan, ``{Celestial Operator
  Products of Gluons and Gravitons},''
\href{http://arxiv.org/abs/1910.07424}{{\ttfamily arXiv:1910.07424 [hep-th]}}.

\bibitem{Ball:2019atb}
A.~Ball, E.~Himwich, S.~A. Narayanan, S.~Pasterski, and A.~Strominger,
  ``{Uplifting AdS$_{3}$/CFT$_{2}$ to flat space holography},''
  \href{http://dx.doi.org/10.1007/JHEP08(2019)168}{{\em JHEP} {\bfseries 08}
  (2019) 168}, \href{http://arxiv.org/abs/1905.09809}{{\ttfamily
  arXiv:1905.09809 [hep-th]}}.

\bibitem{Barnich:2011mi}
G.~Barnich and C.~Troessaert, ``{BMS charge algebra},''
  \href{http://dx.doi.org/10.1007/JHEP12(2011)105}{{\em JHEP} {\bfseries 12}
  (2011) 105},
\href{http://arxiv.org/abs/1106.0213}{{\ttfamily arXiv:1106.0213 [hep-th]}}.

\bibitem{Costa:2011mg}
M.~S. Costa, J.~Penedones, D.~Poland, and S.~Rychkov, ``{Spinning Conformal
  Correlators},'' \href{http://dx.doi.org/10.1007/JHEP11(2011)071}{{\em JHEP}
  {\bfseries 11} (2011) 071},
\href{http://arxiv.org/abs/1107.3554}{{\ttfamily arXiv:1107.3554 [hep-th]}}.

\bibitem{Costa:2011dw}
M.~S. Costa, J.~Penedones, D.~Poland, and S.~Rychkov, ``{Spinning Conformal
  Blocks},'' \href{http://dx.doi.org/10.1007/JHEP11(2011)154}{{\em JHEP}
  {\bfseries 11} (2011) 154},
\href{http://arxiv.org/abs/1109.6321}{{\ttfamily arXiv:1109.6321 [hep-th]}}.

\bibitem{SimmonsDuffin:2012uy}
D.~Simmons-Duffin, ``{Projectors, Shadows, and Conformal Blocks},''
  \href{http://dx.doi.org/10.1007/JHEP04(2014)146}{{\em JHEP} {\bfseries 04}
  (2014) 146},
\href{http://arxiv.org/abs/1204.3894}{{\ttfamily arXiv:1204.3894 [hep-th]}}.

\bibitem{Costa:2014kfa}
M.~S. Costa, V.~Gonçalves, and J.~Penedones, ``{Spinning AdS Propagators},''
  \href{http://dx.doi.org/10.1007/JHEP09(2014)064}{{\em JHEP} {\bfseries 09}
  (2014) 064},
\href{http://arxiv.org/abs/1404.5625}{{\ttfamily arXiv:1404.5625 [hep-th]}}.

\bibitem{Ashtekar:1987tt}
A.~Ashtekar, {\em {Asymptotic Quantization: Based on 1984 Naples Lectures}}.
\newblock
1987.
\newblock

\bibitem{Crnkovic:1986ex}
C.~{Crnkovic} and E.~{Witten}, ``{Covariant description of canonical formalism
  in geometrical theories},'' in {\em Three Hundred Years of Gravitation},
  pp.~676--684.
\newblock {S.~W. Hawking} and { W. Israel}, 1987.

\bibitem{Lee:1990nz}
J.~Lee and R.~M. Wald, ``{Local symmetries and constraints},''
\href{http://dx.doi.org/10.1063/1.528801}{{\em J. Math. Phys.} {\bfseries 31}
  (1990) 725--743}.

\bibitem{Wald:1999wa}
R.~M. Wald and A.~Zoupas, ``{A General definition of `conserved quantities' in
  general relativity and other theories of gravity},''
  \href{http://dx.doi.org/10.1103/PhysRevD.61.084027}{{\em Phys. Rev.}
  {\bfseries D61} (2000) 084027},
\href{http://arxiv.org/abs/gr-qc/9911095}{{\ttfamily arXiv:gr-qc/9911095
  [gr-qc]}}.

\bibitem{upcoming3}
L.~Donnay, S.~Pasterski, and A.~Puhm, ``{Celestial Primaries and Their
  Memories},''
{\em in preparation} .

\bibitem{upcoming2}
L.~Donnay, S.~Pasterski, and A.~Puhm, ``{Conformal Soft Theorems without
  Conformal Goldstones},''
{\em to appear} .

\bibitem{He:2014cra}
T.~He, P.~Mitra, A.~P. Porfyriadis, and A.~Strominger, ``{New Symmetries of
  Massless QED},'' \href{http://dx.doi.org/10.1007/JHEP10(2014)112}{{\em JHEP}
  {\bfseries 10} (2014) 112},
\href{http://arxiv.org/abs/1407.3789}{{\ttfamily arXiv:1407.3789 [hep-th]}}.

\bibitem{He:2014laa}
T.~He, V.~Lysov, P.~Mitra, and A.~Strominger, ``{BMS supertranslations and
  Weinberg'€™s soft graviton theorem},''
  \href{http://dx.doi.org/10.1007/JHEP05(2015)151}{{\em JHEP} {\bfseries 05}
  (2015) 151},
\href{http://arxiv.org/abs/1401.7026}{{\ttfamily arXiv:1401.7026 [hep-th]}}.

\bibitem{Avery:2015gxa}
S.~G. Avery and B.~U.~W. Schwab, ``{Burg-Metzner-Sachs symmetry, string theory,
  and soft theorems},''
  \href{http://dx.doi.org/10.1103/PhysRevD.93.026003}{{\em Phys.\ Rev.\ D}
  {\bfseries 93} (2016) 026003},
  \href{http://arxiv.org/abs/1506.05789}{{\ttfamily arXiv:1506.05789
  [hep-th]}}.

\bibitem{Campiglia:2016efb}
M.~Campiglia and A.~Laddha, ``{Sub-subleading soft gravitons and large
  diffeomorphisms},'' \href{http://dx.doi.org/10.1007/JHEP01(2017)036}{{\em
  JHEP} {\bfseries 01} (2017) 036},
\href{http://arxiv.org/abs/1608.00685}{{\ttfamily arXiv:1608.00685 [gr-qc]}}.

\bibitem{Himwich:2019qmj}
E.~Himwich, Z.~Mirzaiyan, and S.~Pasterski, ``{A Note on the Subleading Soft
  Graviton},''
\href{http://arxiv.org/abs/1902.01840}{{\ttfamily arXiv:1902.01840 [hep-th]}}.

\bibitem{Compere:2019odm}
G.~Comp\`ere, ``{Infinite towers of supertranslation and superrotation
  memories},'' \href{http://dx.doi.org/10.1103/PhysRevLett.123.021101}{{\em
  Phys. Rev. Lett.} {\bfseries 123} no.~2, (2019) 021101},
  \href{http://arxiv.org/abs/1904.00280}{{\ttfamily arXiv:1904.00280 [gr-qc]}}.

\bibitem{Barnich:2013axa}
G.~Barnich and C.~Troessaert, ``{Comments on holographic current algebras and
  asymptotically flat four dimensional spacetimes at null infinity},''
  \href{http://dx.doi.org/10.1007/JHEP11(2013)003}{{\em JHEP} {\bfseries 11}
  (2013) 003}, \href{http://arxiv.org/abs/1309.0794}{{\ttfamily arXiv:1309.0794
  [hep-th]}}.

\bibitem{Barnich:2017ubf}
G.~Barnich, ``{Centrally extended BMS4 Lie algebroid},''
  \href{http://dx.doi.org/10.1007/JHEP06(2017)007}{{\em JHEP} {\bfseries 06}
  (2017) 007}, \href{http://arxiv.org/abs/1703.08704}{{\ttfamily
  arXiv:1703.08704 [hep-th]}}.

\bibitem{Strominger:2016wns}
A.~Strominger and A.~Zhiboedov, ``{Superrotations and Black Hole Pair
  Creation},'' \href{http://dx.doi.org/10.1088/1361-6382/aa5b5f}{{\em Class.
  Quant. Grav.} {\bfseries 34} (2017) 064002},
\href{http://arxiv.org/abs/1610.00639}{{\ttfamily arXiv:1610.00639 [hep-th]}}.

\bibitem{Compere:2018aar}
G.~Comp\`ere and A.~Fiorucci, ``{Advanced Lectures on General Relativity},''
  \href{http://arxiv.org/abs/1801.07064}{{\ttfamily arXiv:1801.07064
  [hep-th]}}.

\bibitem{Iyer:1994ys}
V.~Iyer and R.~M. Wald, ``{Some properties of Noether charge and a proposal for
  dynamical black hole entropy},''
  \href{http://dx.doi.org/10.1103/PhysRevD.50.846}{{\em Phys. Rev. D}
  {\bfseries 50} (1994) 846--864},
  \href{http://arxiv.org/abs/gr-qc/9403028}{{\ttfamily arXiv:gr-qc/9403028}}.

\bibitem{Distler:2018rwu}
J.~Distler, R.~Flauger, and B.~Horn, ``{Double-soft graviton amplitudes and the
  extended BMS charge algebra},''
  \href{http://dx.doi.org/10.1007/JHEP08(2019)021}{{\em JHEP} {\bfseries 08}
  (2019) 021}, \href{http://arxiv.org/abs/1808.09965}{{\ttfamily
  arXiv:1808.09965 [hep-th]}}.

\bibitem{Dolan:2011dv}
F.~A. Dolan and H.~Osborn, ``{Conformal Partial Waves: Further Mathematical
  Results},''
\href{http://arxiv.org/abs/1108.6194}{{\ttfamily arXiv:1108.6194 [hep-th]}}.

\bibitem{Compere:2020lrt}
G.~Comp\`ere, A.~Fiorucci, and R.~Ruzziconi, ``{The $\Lambda$-BMS$_4$ Charge
  Algebra},'' \href{http://arxiv.org/abs/2004.10769}{{\ttfamily
  arXiv:2004.10769 [hep-th]}}.

\end{thebibliography}\endgroup

\end{document}